\documentclass[11pt,a4paper]{article}
\usepackage{jheppub}
\usepackage[T1]{fontenc} 
\hypersetup{
	colorlinks=true,
	linkcolor=blue,
	filecolor=blue,      
	urlcolor=blue,
	citecolor=blue
}
\usepackage{amsmath,amssymb,amsthm}
\usepackage{mathtools}
\usepackage{physics}
\usepackage{enumitem}
\usepackage{caption,subcaption}
\usepackage{hyperref}
\usepackage{cleveref}
\usepackage{xcolor}
\usepackage{extarrows}
\usepackage{empheq}
\usepackage{scalerel}
\usepackage{multirow}
\usepackage{makecell}
\usepackage{pgfplots}
\pgfplotsset{compat=1.18}
\usepackage{upgreek}
\usepackage{graphicx} 
\usepackage[compat=1.0.0]{tikz-feynman}
\usepackage{tcolorbox}
\usepackage{cancel}
\usepackage{soul}
\usepackage{color}
\usepackage{listings}
\usepackage{exercise}
\usepackage{cases}
\usepackage{todonotes}
\usepackage{fancyhdr}
\usepackage{mdframed}
\usetikzlibrary{arrows.meta,calc,decorations.markings}

\usetikzlibrary{calc}
\mdfdefinestyle{exampledefault}{
	outerlinewidth=5pt,innerlinewidth=0pt,
	outerlinecolor=red,roundcorner=5pt
}
\newcommand{\tauacc}{\tau_{\scaleto{\rm acc}{2.5pt}}}
\newcommand{\taureg}{\tau_{\scaleto{\rm reg}{4pt}}}
\newcommand{\gammaterm}{\gamma_{\scaleto{\rm term}{3.8pt}}}
\newcommand{\gammaeff}{\gamma_{\scaleto{\rm eff}{4.5pt}}}
\newcommand{\betagw}{\beta_{\scaleto{\rm GW}{4pt}}}
\newcommand{\muqp}{\mu_{\scaleto{\cancel{\rm PQ}}{5pt}}}
\newcommand{\vqp}{V_{\scaleto{\cancel{\rm PQ}}{5pt}}}
\newcommand{\tbbn}{T_{\scaleto{\rm BBN}{4pt}}}
\newcommand{\epsdm}{\varepsilon_{\scaleto{\rm DM}{4pt}}}
\newcommand{\epsdr}{\varepsilon_{\scaleto{\rm DR}{4pt}}}
\newcommand{\rhodr}{\rho_{\scaleto{\rm DR}{4pt}}}
\newcommand{\rhodm}{\rho_{\scaleto{\rm DM}{4pt}}}
\newcommand{\Nc}{N_{\rm c}}
\newcommand{\tcr}{T_{\rm cr}}
\newcommand{\tnuc}{T_{\rm n}}
\newcommand{\triplediagram}[2][0.38]{
  \vcenter{\hbox{
    \begin{tikzpicture}[
      scale=#1,
      transform shape,
      line cap=round,
      line join=round
    ]
\draw (0,0) circle[radius=1.05];
\node [scale=2] at (0,0) {\Large $#2$};
\coordinate (A) at (90:3.0);
\coordinate (B) at (210:3.0);
\coordinate (C) at (330:3.0);
\draw (90:1.05)  -- (90:2.7);
\draw (210:1.05) -- (210:2.7);
\draw (330:1.05) -- (330:2.7);
\foreach \ang in {114,138,162,186}{
  \fill (\ang:1.62) circle[radius=0.10];
}
\foreach \ang in {234,258,282,306}{
  \fill (\ang:1.62) circle[radius=0.10];
}
\foreach \ang in {354,378,402,426}{
  \fill (\ang:1.62) circle[radius=0.10];
}
\foreach \p/\ang in {A/45,B/175,C/285}{
  \draw (\p) circle[radius=0.27];
  \draw
    ($(\p)+(\ang:0.27)$) --
    ($(\p)+({\ang+180}:0.27)$);
  \draw
    ($(\p)+({\ang+90}:0.27)$) --
    ($(\p)+({\ang+270}:0.27)$);
}
\end{tikzpicture}
  }}
}
\newcommand{\cuttriplediagram}[2][0.38]{
  \vcenter{\hbox{
    \begin{tikzpicture}[
      scale=#1,
      transform shape,
      line cap=round,
      line join=round
    ]
      \draw (0,0) circle[radius=1.05];
      \node [scale=2] at (0,0) {\Large $#2$};
\draw[blue]
  (-0.45,1.90) -- (0.45,2.10);
\coordinate (A) at (90:3.0);
\coordinate (B) at (210:3.0);
\coordinate (C) at (330:3.0);
\draw (90:1.05)  -- (90:2.7);
\draw (210:1.05) -- (210:2.7);
\draw (330:1.05) -- (330:2.7);
\foreach \ang in {114,138,162,186}{
  \fill (\ang:1.62) circle[radius=0.10];
}
\foreach \ang in {234,258,282,306}{
  \fill (\ang:1.62) circle[radius=0.10];
}
\foreach \ang in {354,378,402,426}{
  \fill (\ang:1.62) circle[radius=0.10];
}
\foreach \p/\ang in {A/45,B/175,C/285}{
  \draw (\p) circle[radius=0.27];
  \draw
    ($(\p)+(\ang:0.27)$) --
    ($(\p)+({\ang+180}:0.27)$);
  \draw
    ($(\p)+({\ang+90}:0.27)$) --
    ($(\p)+({\ang+270}:0.27)$);
}
\end{tikzpicture}
  }}
}
\title{Abundant production of scalars and axions from phase transition bubble expansion}
\author[\textsuperscript{\fontsize{8}{9}\selectfont$\mathbb{W}$}]{Isabel Garcia Garcia,}
\author[\textsuperscript{\fontsize{10.5}{11}\selectfont$\upvarphi$}]{Gaurang Ramakant Kane,}
\author[\textsuperscript{\fontsize{10.5}{11}\selectfont$\upvarphi$}]{John March-Russell,}
\author[\textsuperscript{\fontsize{8}{9}\selectfont$\mathbb{W}$}]{and Andrea Paolini}
\affiliation[\textsuperscript{$\mathbb{W}$}]{Department of Physics, University of Washington, Seattle, WA 98195, USA}
\affiliation[\textsuperscript{\fontsize{9}{10}\selectfont$\upvarphi$}]{Rudolf Peierls Centre for Theoretical Physics, University of Oxford, Parks Road, Oxford, OX1 3PU, United Kingdom}
\emailAdd{isabelgg@uw.edu}
\emailAdd{gaurang.kane@physics.ox.ac.uk}
\emailAdd{john.march-russell@physics.ox.ac.uk}
\emailAdd{andrepao@uw.edu}

\abstract{We revisit aspects of particle production during cosmological first-order phase transitions, and show that the expansion of true-vacuum bubbles can be a copious source of particle production. Specifically, for a massive spin-0 field linearly coupled to the bubble profile, spherical bubbles expanding even at \emph{constant} radial velocity efficiently produce particles until the local-rest-frame radius of curvature of the bubble wall exceeds the particle Compton wavelength -- contrary to the expectation that walls moving at constant speed cannot radiate. We compute the momentum spectrum of the produced particles for both accelerated and constant-velocity expansion histories, identify the regimes of coherent and incoherent production, quantify the validity of the perturbative treatment, and show that this mechanism can parametrically dominate other production processes of feebly coupled particles, including freeze-in. Our results apply to many models of light, feebly coupled particles studied in the literature, providing new sources of dark radiation and dark matter. In particular, a first-order deconfinement-confinement transition in a hidden Yang-Mills sector can abundantly produce axion-like particles, leading to dark radiation and/or dark matter signatures over large regions of parameter space.}

\begin{document}
\maketitle
\section{Introduction}
\label{sec:introduction}
First-order phase transitions (FOPTs) are ubiquitous in both the everyday world and in many motivated extensions of the Standard Model. In the context of early universe cosmology, a FOPT can have far-reaching consequences, from baryogenesis~\cite{Kuzmin:1985mm, Shaposhnikov:1987tw, Cohen:1993nk}, to relics surviving until the present day~\cite{Asadi:2021pwo,Gouttenoire:2023roe}, to the generation of a stochastic gravitational wave background~\cite{Witten:1984rs, Hogan:1986qda, Kosowsky:1991ua, Kosowsky:1992rz, Kamionkowski:1993fg}. The latter possibility is particularly exciting given the current and near-future sensitivities of gravitational wave observatories in various frequency bands. These phase transitions proceed by nucleation of bubbles of the true vacuum, as first discussed by Langer \cite{Langer:1969bc} in the context of condensed matter systems and by Coleman and Callan \cite{Coleman:1977py, Callan:1977pt} and Linde \cite{Linde:1981zj} for relativistic field theories at zero and non-zero temperature, respectively. Depending on whether the transition is dominated by quantum fluctuations or thermal fluctuations, the bubbles have, respectively, either an $O(4)$-symmetric \cite{Coleman:1977th} or an $O(3)$-symmetric \cite{Linde:1981zj} nucleation structure.\footnote{These are the two standard FOPT bubble ansatz forms. In the presence of background fields \cite{Hassan:2024nbl}, or defects (``dirt''), see e.g. \cite{Agrawal:2022hnf,Chatrchyan:2025uar}, the symmetry group can be different. We focus on spherically symmetric solutions.} Following nucleation, bubbles larger than a certain critical size expand due to the pressure difference across the bubble wall. In the absence of significant friction increasing with wall velocity, the bubble wall achieves constant proper acceleration, which continues until the bubbles collide. This scenario is referred to as the \emph{runaway} case \cite{Bodeker:2009qy}. On the other hand, in the presence of a significant velocity-dependent friction due, for instance, to the interaction of the fields that make up the bubble with a thermal plasma, the bubbles can achieve terminal velocities \cite{Turok:1992jp, Turner:1992tz, Moore:1995si, Azatov:2020ufh, Gouttenoire:2021kjv, GarciaGarcia:2022yqb, GarciaGarcia:2024dfx, Ai:2024btx}.

That accelerating bubble walls produce particles has long been recognized. For walls moving at constant speed, however, it is often argued that no production occurs: production is a local process, so one may go to the local rest frame of the wall, in which the source is time-independent, and conclude that no particles are produced~\cite{Shakya:2023kjf, Giudice:2024tcp}. We show that this argument is incomplete: although every point on the wall admits a local rest frame, the curvature of the wall prevents any single inertial frame from rendering static an entire radiation formation region. Through explicit calculations, we show that for light particles significant production can occur during the constant wall speed expansion epoch. Specifically, expansion of spherical bubbles at constant radial velocity efficiently produces spin-0 particles until the \emph{local-rest-frame radius of curvature of the wall  becomes larger than $m^{-1}$.} For spherical terminal-velocity bubbles this means that production does not switch off until the global FRW-frame radius, $R$, of the bubble grows to $R\sim \gamma/m$, where $\gamma$ is the local Lorentz factor of the wall.  So for light particles, the epoch of efficient production can be long-lasting even for walls expanding at a terminal velocity. We compute the momentum spectrum of produced particles for a range of wall trajectories, including constant proper acceleration, constant radial velocity, and a mixed case in which an accelerated wall reaches terminal velocity before collision, and consider both quasi-relativistic and ultra-relativistic expansion.  We find that \emph{both incoherent and coherent} production are possible depending on the bubble size and the produced particle mass, the particle Compton wavelength, $m^{-1}$, being the scale at which production switches from coherent to incoherent.  In the coherently enhanced case the production rate depends on the square of the bubble-wall area, at least until the bubble becomes sufficiently large. In the incoherent case, which might better be called patch-coherent, the production is coherent on small length scales of order the Compton wavelength $m^{-1}$ across the wall, but different patches add incoherently, giving a production scaling as the total wall area.  We also show that this mechanism of particle production can parametrically dominate other production processes of feebly coupled particles, including thermal freeze-in \cite{Hall:2009bx,McDonald:2001vt}.

Following the expansion phase, the bubbles collide, subsequently completing the FOPT. During collision, light particles (also heavy particles, due to altered kinematics) can be produced, a possibility that has been investigated previously~\cite{Watkins:1991zt,Falkowski:2012fb, Shakya:2023kjf, Giudice:2024tcp,Ghoshal:2026pew, An:2026sdu}.\footnote{Reliable estimates of this collision-epoch particle production require detailed numerical simulation of the full multi-bubble field dynamics. Of course, bubble wall collisions are also associated with another crucial aspect of cosmological FOPTs: the emission of gravitational waves \cite{Kosowsky:1991ua, Kosowsky:1992rz, Kosowsky:1992vn, Kamionkowski:1993fg}. In the last decade, with the prospect of near-future detectors sensitive to transitions near the electroweak scale, there has been an increased interest in and study of FOPTs in a variety of beyond-the-Standard-Model scenarios. These studies use both analytical and numerical techniques with the aim of quantifying the gravitational wave spectrum~\cite{Grojean:2006bp, Caprini:2007xq, Hindmarsh:2013xza, Hindmarsh:2015qta, Hindmarsh:2019phv}. An effect often overlooked in these studies is particle production --- a channel into which bubble energy is dumped, reducing the walls' kinetic energy and hence potentially modifying the gravitational wave signatures. Quantifying this back-reaction on the bubble dynamics requires separate dedicated study, and we defer it to future work.} Here we instead focus on the largely overlooked possibility of particle production during bubble expansion, finding that significant production is possible, at least for sufficiently light particles. For bubbles that are well-separated during nucleation, the expansionary production is a single bubble process and is calculable without resorting to numerical simulations.

We emphasize that our focus is on the case where the production from different bubbles does not interfere, so we treat the total expansion-epoch production within a Hubble patch as an incoherent sum of single bubble production processes with bubbles well separated across the Hubble patch. Of course, a single-bubble treatment is not appropriate immediately before or during collision. This situation, in which nucleated bubbles are well separated within a Hubble patch, is characteristic of supercooled phase transitions. For more on the relation between supercooling, the inverse of the duration of the transition until completion $\betagw\sim R_{*}^{-1}$ ($R_{*}$ being the mean radius at the collision epoch), and gravitational wave signals in various models, including FOPTs in strongly coupled theories, see \cite{Turner:1992tz,Hindmarsh:2019phv, GarciaGarcia:2015fol,Gouttenoire:2023roe, Azatov:2020nbe, Agrawal:2025wvf} and references therein.

In the second half of this work, contained in section \ref{sec:alp_pheno_case}, we consider particle production due to bubble wall expansion in the context of a simple and quite natural model: a hidden-sector Yang-Mills theory undergoing a deconfinement-confinement phase transition in the presence of an axion-like particle (ALP) coupled  via the usual $a G {\tilde G}$ term. If the final CP- and P-violating $\theta$-angle in the true-vacuum confined phase YM theory is \emph{not} entirely cancelled, then the physical ALP excitation linearly couples to a $G {\tilde G}$ expectation value that changes across the deconfined-confined bubble wall, thereby leading to particle production during expansion. Such an uncancelled residual $\theta$ angle occurs when there is an additional not-too-small ``axion-quality problem'' potential (which violates by a small amount the axion shift symmetry), which is to be expected for many ALPs \cite{Dine:1986bg, Kamionkowski:1992mf, Barr:1992qq,Holman:1992us}. Indeed, in the QCD axion case this is a feature that we have to work hard to forbid down to acceptably tiny levels! We show that, depending on model parameters, ALPs are readily produced in numbers sufficient to act as dark radiation or dark matter over large regions of parameter space, with warm dark matter in a transition region.

This paper is organized as follows: In section \ref{subsec:set-up}, we give a general treatment of spin-0 particle production due to a linear coupling. In section \ref{sec:bubble_nucleation} we consider, for completeness, production during bubble nucleation, finding it is negligible compared to that during the expansion epoch (apart from an extreme region of parameter space).  In sections \ref{subsec:constant_velocity_and_the_regulator} and \ref{subsec:constant_proper_acceleration} we consider different bubble expansion trajectories and study the energy spectrum of the produced particles.  Section \ref{subsec:cessation} gives a quantitative analysis of when production ceases in the constant velocity case, while section \ref{sec:hard_cutoff} discusses aspects of the patching procedure for more complicated trajectories, and section \ref{sec:finite_wall_thickness} the changes resulting from finite bubble wall thickness effects.  Section \ref{sec:alp_pheno_case} studies a simple model of an axion-like particle coupled to a dark sector $SU(\Nc)$ confining gauge theory.  We compare the number of particles produced from expansion to that produced by the freeze-in process, and study the parameter space where our mechanism leads to significant dark radiation and dark matter production. In section \ref{sec:back-reaction} we estimate the back-reaction effects on the $SU(\Nc)$ bubble wall due to axion production and present an initial partial discussion of whether a terminal-velocity epoch can result.  Section \ref{sec:discussion} concludes with a summary and directions for future work. Appendix \ref{appendix:weak_coupling} quantifies what it means for the source to be weakly coupled when production is coherent, as it is for us in large parts of parameter space, and appendix \ref{appendix:transition_dynamics} summarizes aspects of thermal FOPT dynamics for our $SU(\Nc)$ confining gauge theory case.

\section{Particle production from expanding bubbles}
\label{sec:Particle production from expanding bubbles}
\subsection{Setup and basic properties of the produced spectrum}
\label{subsec:set-up}
Our starting point is a massive spin-0 field $\varphi$ linearly coupled to a source $J$ describing the true-vacuum bubble:
\begin{equation}
    \mathcal{L}=\dfrac{1}{2}\partial_{\mu}\varphi\partial^{\mu}\varphi-\dfrac{1}{2}m^{2}\varphi^{2}+J\varphi+\cdots~~.
    \label{eqn:general_lagrangian}
\end{equation}
Here we have absorbed the coupling strength between the sector driving the first-order phase transition and the field $\varphi$ into the definition of $J$. As we discuss in section \ref{sec:alp_pheno_case}, this linear coupling is well motivated in a variety of scenarios: for example, a light axion-like particle interacting via $(a/f_a) \Tr (G\widetilde{G})$ where $G$ is the field strength of a Yang-Mills theory undergoing a thermal confining phase transition. In this case the source strength is proportional to $\sin\theta_{\rm eff} \Lambda^4/f_a$ where $\Lambda$ is the YM dynamical scale, and $\theta_{\rm eff}$ is the residual uncancelled CP-violating angle due to the extra quality problem potential. We assume the source is weak enough that the production rate can be computed perturbatively and, relatedly, that any self-interactions of $\varphi$ are too feeble to alter our results. Below, and in appendix \ref{appendix:weak_coupling}, we quantify the meaning of ``weak coupling'' and show that the resulting bound is easy to satisfy in many phenomenological implementations of this particle production mechanism.

Let us first discuss the key physical properties of the source $J$ in equation \eqref{eqn:general_lagrangian}. Bubbles at nucleation are conventionally classified as thick-walled or thin-walled (see, e.g., \cite{Weinberg:2012pjx} for a pedagogical discussion). In the thick-wall case, usually, there is no parametric separation between the initial critical size of the bubble and its wall thickness, whereas, in the second case, there are two different scales that set these two lengths with the wall thickness being parametrically smaller than the critical bubble radius. As the bubbles expand, however, even the thick-wall case becomes well approximated by a thin wall --- increasingly so for relativistic walls, whose thickness is Lorentz-contracted. Hence, in our treatment, we focus on the (infinitely) thin wall case, although we will discuss the effect of finite wall thickness in section \ref{sec:finite_wall_thickness}. 
We also assume a spherically symmetric source
\begin{equation}
    J(t,\vec{x})=J_{0}f(t,r)~,
    \label{eqn:general_spherical_source}
\end{equation}
where $J_{0}$ is a constant, parametrizing the strength of the source. For a spin-0 field, no angular momentum selection rule forbids production from an exactly spherical bubble --- an important difference from massless spin-1 or spin-2 fields. A second crucial difference is that, unlike massless higher-spin fields, our spin-0 field is consistently coupled to a source that is \emph{not} conserved, and we find that this leads to a large enhancement of the production rate.

Standard perturbation theory (see, e.g., \cite{Peskin:1995ev}) gives the energy spectrum as a function of the magnitude of the particle momentum $k$ as
\begin{equation}
    \dfrac{d\langle E\rangle }{dk}=\dfrac{k^{2}}{4\pi^{2}}\left\vert \widetilde{J}(\omega_{k},k)\right\vert^{2},\qquad\omega_{k}=\sqrt{k^{2}+m^{2}}~~,
    \label{eqn:energy_spectrum_peskin}
\end{equation}
where we have only kept the leading, quadratic in $J$, contribution.  $\widetilde{J}(\omega_{k},k)$ is the Fourier transform (FT) of the source $J(t,\vec{x})$ constrained to satisfy the on-shell condition.

Equation \eqref{eqn:energy_spectrum_peskin} holds for a free field. If $\varphi$ has a non-trivial potential, and hence self-interactions, the leading-order treatment remains valid provided the self-coupling is weak. For a potential $V(\varphi)$, one must ensure that $\varphi$ is stabilized at its minimum and that the term $J\varphi$ does not deform the potential at leading order, i.e., that fluctuations about this minimum can be treated as weakly interacting $\varphi$ quanta with $J$ coupled to these fluctuations. In this case, the mass $m$ comes from the curvature of the potential around the minimum and for an $n^{\rm th}-$order scalar self-interaction $(\lambda_n/n!) \varphi^n$, with $n\geq3$, the linear response around this minimum is reliable when $\lambda_n\varphi_{\rm cl}^n/n!\ll \varphi_{\rm cl}^2 m^2$, which upon using $\varphi_{\rm cl}\sim J_0/m^2$ gives the condition
\begin{equation}
2\frac{\lambda_n}{n!}\frac{J_0^{n-2}}{m^{2n-2}}\ll 1~.
\label{eq:weak_source}
\end{equation}
This constraint is easily met in implementations of our model; see section \ref{sec:alp_pheno_case} for the axion case where, of course, parametrically tiny self-interactions, e.g., $\lambda_4 \sim \muqp^4/f_a^4 \ll 1$, are commonplace. (For more discussion of this condition see appendix \ref{appendix:weak_coupling}.)

Once we make sure that the self-interactions are weak around the minimum we can use the free field result. In addition, we are considering a source that will at the completion of the transition have a uniform (true) vacuum expectation value (VEV) throughout space. One might worry that the energy computed for $\varphi$ fails to account for the VEV shift the field undergoes at the end of the transition, i.e.~that we are expanding around the wrong vacuum when counting $\varphi$-particles inside the bubble. However, for the energy and number of on-shell particles, this shift is irrelevant.  To prove this, define
\begin{equation} 
    \phi(x)=\varphi(x)-J(x)/m^2~~,
    \label{eqn:phivarphi}
\end{equation}
so that equation \eqref{eqn:general_lagrangian} is expanded around the minimum $J_0/m^2$ of the potential inside the bubble, and around $0$ outside. This will automatically incorporate the correct expansion around the homogeneous VEV when the transition is over. Substituting equation \eqref{eqn:phivarphi} into equation \eqref{eqn:general_lagrangian},
\begin{equation}
    \mathcal{L}(\phi)=\dfrac{1}{2}\partial_{\mu}\phi\partial^{\mu}\phi-\dfrac{1}{2}m^{2}\phi^{2}-\frac{\square J}{m^2}\phi + (\text{vacuum energy})~~,
\end{equation}
so the shifted field has source $\square J/m^2$. Since we are concerned only with the field generated by the on-shell Fourier modes of the source, i.e. with $p^2=m^2$,  we conclude that the on-shell modes generated in $\varphi$ and $\phi$ are the same, and a treatment following \cite{Peskin:1995ev} is still valid. What will vary between the two fields is the amount of energy they have partitioned between the Yukawa cloud and the vacuum energy of the system, and for such purposes one should use the shifted field $\phi$ instead of $\varphi$. The VEV shift will also cause a general shift of the couplings; in particular, the mass receives corrections 
\begin{equation}
    \frac{\Delta m^{2}}{m^2}=\sum_{n=3}^\infty \frac{\lambda_n}{(n-2)!}\frac{J_0^{n-2}}{m^{2n-2}}~~.
    \label{eq:mass_correction}
\end{equation}
Hence, in addition to imposing equation \eqref{eq:weak_source}, one must check that this series converges and yields small corrections to the mass, as we verify for the model of section \ref{sec:alp_pheno_case}.

As we are considering spherically symmetric sources, the Fourier transform takes the form
\begin{eqnarray}
   \widetilde{J}(\omega_k,k)
   &=&\dfrac{4\pi J_{0}}{k}\int_{-\infty}^{\infty} dt~e^{i\omega_k t}\int_{0}^{\infty} dr~rf(t,r)\sin(kr) ~~.
\label{eqn:spherically_sym_fourier}
\end{eqnarray}
To go further, we need to specify the functional form of $f(t,r)$ that encapsulates both the profile details, such as the wall thickness, and the bubble wall spacetime trajectory. First, however, let us address the particle production that arises from the initial epoch of bubble \emph{nucleation}. We will see that in almost all reasonable cases the
production during expansion parametrically dominates that from nucleation.  Readers most interested in production from expansion can jump to section \ref{subsec:constant_velocity_and_the_regulator}.

\subsection{Particle production from bubble nucleation}
\label{sec:bubble_nucleation}
Before turning to expansion, we consider particle production during the nucleation of true-vacuum bubbles, following the procedure of~\cite{Rubakov:1984pa}, with the only novelty that our particles couple linearly, rather than quadratically, to the bubble wall. Our results will allow us to establish that particle production due to nucleation is sub-leading to that during bubble expansion, excepting the extreme case where $m R_0\gtrsim 1$ ($R_0$ is the bubble radius at nucleation).

Following \cite{Rubakov:1984pa}, we take the field $\varphi$ during the transition to evolve according to the Euclidean equation of motion with the Euclidean continuation of the source
\begin{equation}
    (-\partial^2_\tau- \nabla^2+ m^2)\varphi(\tau,\vec{x})=J(\tau,\vec{x})~~.
    \label{eqn:Euclidean_equation_of_motion}
\end{equation}
The solution to this equation can be separated into the homogeneous solution, $\varphi_{0}$, and a particular solution. The homogeneous piece is just the Euclidean version of the free scalar field expansion, with the substitution $t=-i\tau$,
\begin{equation}
    \varphi_0(\tau,\vec{x})= \int \frac{d^3 k}{(2\pi)^3}\frac{1}{\sqrt{2 \omega_{k}}} \left(b_{\vec{k}}\, e^{-\omega_{k}\tau+ i\vec{k}\cdot \vec{x}}+ b^\dagger_{\vec{k}}\, e^{\omega_{k}\tau- i\vec{k}\cdot \vec{x}}\right), \quad \omega_{k}=\sqrt{|\vec{k}|^2+m^2} \ ,
\end{equation}
where $b^\dagger_{\vec{k}}$ and $b_{\vec{k}}$ are the usual creation and annihilation operators defined in the false vacuum. Using the Green's function $G_{\scaleto{\rm E}{4pt}}$ for the differential operator of equation \eqref{eqn:Euclidean_equation_of_motion} 
\begin{equation}
    G_{\scaleto{\rm E}{4pt}} (\tau, \vec x)= \int \frac{d^3 k}{(2\pi)^3} \frac{1}{2\omega_{k}} \Big(\Theta(\tau) e^{-\omega_{k}\tau- i\vec{k}\cdot\vec{x}}+ \Theta(-\tau) e^{\omega_{k}\tau+ i\vec{k}\cdot\vec{x}}\Big)~~,
    \label{eqn:Euclidean_green_function}
\end{equation}
together with the homogeneous solution, we can write the complete solution of  equation \eqref{eqn:Euclidean_equation_of_motion} as
\begin{equation}
    \varphi(x)=\int \dfrac{d^{3}k}{(2\pi)^{3}}\dfrac{1}{\sqrt{2\omega_{k}}}\left(\alpha_{\vec{k}}e^{\omega_{k}\tau+i\vec{k}\cdot\vec{x}}+\beta_{\vec{k}}^{\dagger} e^{-\omega_{k}\tau-i\vec{k}\cdot\vec{x}}\right)
    \label{eqn:scalar_field_euclidean}
\end{equation}
where
\begin{align}
    \alpha_{\vec{k}} (\tau) &= b_{\vec{k}}+ \frac{1}{\sqrt{2 \omega_{k}}}\int d\tau'\int d^3y\, \Theta(\tau'-\tau)J(y)e^{-\omega_{k}\tau'- i\vec{k}\cdot \vec{y}}~~,\nonumber\\
    \beta^\dagger_{\vec{k}} (\tau) &= b^\dagger_{\vec{k}}+\frac{1}{\sqrt{2 \omega_{k}}}\int d\tau'\int d^3y\, \Theta(\tau-\tau')J(y)e^{\omega_{k}\tau'+ i\vec{k}\cdot \vec{y}}~~.
    \label{creann}
\end{align}
Here, $\alpha_{\vec{k}}$ and $\beta^\dagger_{\vec{k}}$ are the annihilation and creation operators in the true vacuum. Notice that $\beta^\dagger_{\vec{k}}$ is not the complex conjugate of $\alpha_{\vec{k}}$. This is characteristic of the Euclidean theory, in which Hermiticity is replaced by reflection positivity, leading to $\varphi^\dagger(-\tau,\vec{x})=\varphi(\tau,\vec{x})$.

We are interested in the energy in the on-shell $\varphi$ particles at bubble nucleation, $\tau=0$. Going to Euclidean signature we need to express the momentum operator in terms of the $\tau$ derivative, $\Pi=\partial_t \varphi=-i\partial_\tau \varphi$. We then find the Hamiltonian operator
\begin{equation}
    H(\tau)= \int \frac{d^3 k}{(2\pi)^3} \frac{\omega}{2}\left(\alpha_{\vec{k}}(\tau) \beta^\dagger_{\vec{k}}(\tau)+ \beta^\dagger_{\vec{k}}(\tau) \alpha_{\vec{k}}(\tau)\right)- \int d^3x J(\tau,\vec{x})\varphi(\tau,\vec{x})~~.
\end{equation}
For definiteness, let us take the nucleating bubble source to be a thin-wall $O(4)$-symmetric bounce solution appropriate for a zero-temperature quantum nucleation. The $O(3)$ symmetric finite-temperature case is analogous, yielding particle production that is likewise negligible compared to expansion, with the same scaling discussed below. Notice that with an $O(4)$-symmetric source, the two source-dependent terms are equal and real. Then the expectation value of the quadratic part of the Hamiltonian evaluated on the vacuum state at $\tau=0$ is
\begin{equation}
    \langle H\rangle= \int \frac{d^3 k}{(2\pi)^3} \frac{1}{2} \left(\int^\infty_0 d\tau\, \int d^3y\, J(\tau,\vec{y})e^{-\omega_{k}\tau+ i\vec{k}\cdot \vec{y}}\right)^2~.
    \label{eqn:hamiltonian_nucleation}
\end{equation}

In the thin-wall case we can approximate the bubble as a step function with radius $R_{0}$ that after nucleation expands with proper acceleration $R_{0}^{-1}$ so
\begin{equation}
    J(\tau,\vec{x})= J_0 \Theta\left(R_{0}^2-\tau^2-r^2\right)~~.
\end{equation}
The procedure is identical to the Minkowski one but with the analytically continued FT. Performing the spatial integral we get
\begin{equation*}
    \widetilde{J}(\omega,k)=\frac{4\pi J_0}{k^3} \int_0^1 dx\, e^{-\omega R_0 x} \left(\sin\left(kR_0\sqrt{1-x^2}\right)- kR_0\sqrt{1-x^2} \cos\left(kR_0\sqrt{1-x^2}\right)\right)~~.
\label{eqn:nucleation_integral_simplification}
\end{equation*}
We can write the oscillating part of the above integral as a series expansion
\begin{equation*}
    \sin(k f(t))-k f(t)\cos(k f(t))
    =\sum_{n=0} \frac{2n(-1)^{n+1}}{(2n+1)!}(k f(t))^{2n+1}~~,
\end{equation*}
and substituting this expansion and performing the integral simplifies the Fourier transform of the source to 
\begin{align}
    \widetilde{J}(\omega,k)
    &=\frac{2\pi^{2}J_{0}}{k^{3}}\frac{(kR_{0})^3}{(\omega R_{0})^2}\sum_{n=0}\frac{(-1)^n}{n!} \left(\frac{(kR_{0})^2}{2\omega R_{0}}\right)^n (\mathrm{I}_{n+2}(\omega R_{0})-\mathrm{L}_{n+2}(\omega R_{0})),
\end{align}
where $\mathrm{I}_{n}$ and $\mathrm{L}_{n}$ are the $n^{\rm th}$ modified Bessel-I function and Struve-L function, respectively. From this expression the total energy emitted by nucleation is
\begin{equation} \label{eq:Enucl}
    \langle E\rangle \Big\vert_{\rm Nuc}\approx \begin{cases}
      \dfrac{4\pi J_{0}^{2}R_{0}^{5}}{15}~,~~~mR_{0}\ll1   \\
        \\
        \dfrac{2\pi J_{0}^{2} R_{0}^{3}}{3m^{2}}~,~~~ mR_{0}\gg 1
    \end{cases}~~.
\end{equation}
These limiting forms have a simple physical interpretation: in the case $mR_0\ll 1$, the production is coherently enhanced across the entire $\sim R_0^2$ surface area of the bubble wall which effectively moves a distance $\sim R_0$ during nucleation.  This leads to an effective production amplitude-squared that is ${\cal O}(R_0^6)$. Since the momentum (and thus energy) of these light particles is set by $1/R_0$ the energy deposition is ${\cal O}(R_0^5)$.  On the other hand, in the heavy particle limit, $mR_0\gg 1$, the production is only coherent over patches on the wall of area $1/m^2$, the contribution of the $R_0^2 m^2$ patches adding incoherently. This leads to the ${\cal O}(R_0^3/m^2)$ behaviour in the $mR_0\gg 1$ limit. We will encounter the same distinction between fully coherent and patch-coherent production in the expansion case.

After nucleation, true vacuum bubbles expand with a trajectory that can include both accelerated and constant-velocity epochs depending on the microphysics governing the transition and the interaction of the bubble walls with the surrounding medium. In Sections~\ref{subsec:constant_velocity_and_the_regulator}-\ref{subsec:constant_proper_acceleration} we show that particle production during expansion dominates that during nucleation, equation \eqref{eq:Enucl}, provided $m R_{0} \ll 1$ and expansion occurs over timescales large compared to $R_{0}$.

\subsection{Coherent and incoherent production from constant velocity expansion}
\label{subsec:constant_velocity_and_the_regulator}
In this subsection, we consider particle production due to bubble walls that expand at \emph{constant} velocity. This scenario deserves attention for two reasons. First, it has previously been argued that spherical bubble walls expanding at constant speed do not radiate particles~\cite{Shakya:2023kjf, Giudice:2024tcp}. As we now show, this conclusion is incomplete: particles can be produced copiously even in this regime. Second, immediately after nucleation, the vacuum bubbles produced during a FOPT begin expanding on a hyperbolic trajectory (with constant proper acceleration), but it is often the case that they asymptote to a terminal velocity due to friction with the surrounding thermal bath of particles \cite{Turok:1992jp, Turner:1992tz, Moore:1995si, Azatov:2020ufh, Gouttenoire:2021kjv, GarciaGarcia:2022yqb, GarciaGarcia:2024dfx, Ai:2024btx}. Thus, understanding particle production at constant velocity is an important ingredient in many cases. As we will see, the total energy of produced particles is \emph{finite} even for eternal expansion.
\begin{figure}[t!]
\centering
\includegraphics[width=0.6\linewidth]{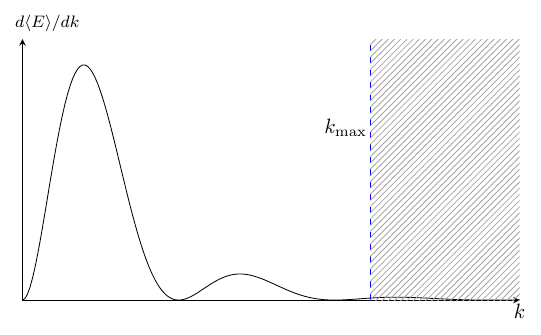}
\caption{A schematic particle spectrum. For $k>k_{\rm max}$ the $\Theta$-function wall approximation fails and finite wall thickness effects must be taken into account. See section \ref{sec:finite_wall_thickness}.}
\label{fig:schematic_version}
\end{figure}

\begin{figure}
\centering
\includegraphics[width=0.7\linewidth]{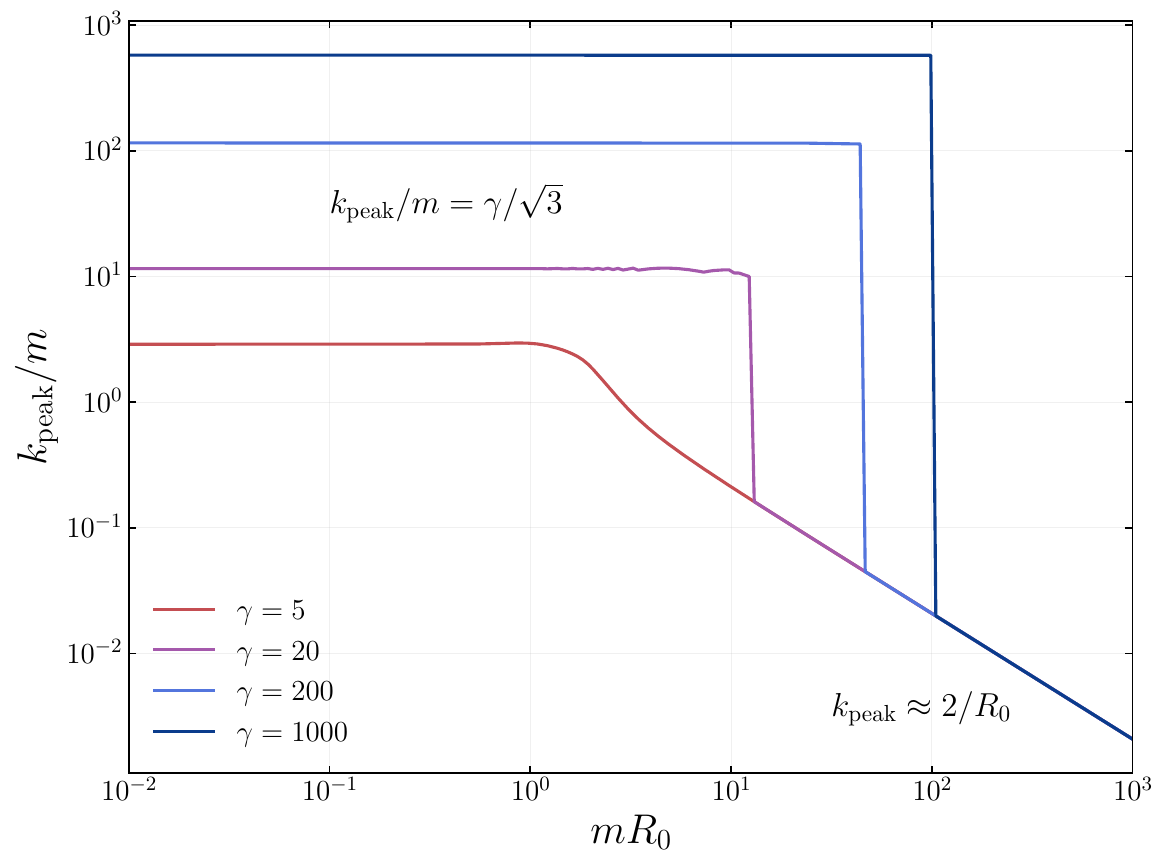}
    \caption{Position, $k_{\rm peak}$, of the dominant peak of the energy spectrum equation~\eqref{eqn:j_mod_squared}. For $mR_{0}\ll \gamma $, $k_{\rm peak}\approx \gamma m/\sqrt{3}$. For $mR_{0}\gg \gamma $, $k_{\rm peak}\approx  2/R_{0}$. The sudden change in $k_{\rm peak}$ curves corresponding to $\gamma=\{20, 200, 1000\}$ is due to a switch-over of the dominant peak.}
    \label{fig:peak_scaling_for_constant_vel}
\end{figure}

Specifically, consider an infinitely thin-wall bubble with initial radius $R(t=0)=R_{0}$ and constant expansion speed $v$.  Thus the source takes the form
\begin{equation}
    J(t,\vec{x})=J_{0} \Theta\left(R_{0}+vt-r\right)\Theta(t)~~,
    \label{eqn:theta_source_const_vel}
\end{equation}
where $\Theta$ is the Heaviside step function. The FT of this is, after an integration by parts,
\begin{equation}
    \widetilde{J}(\omega,k)=\dfrac{4\pi J_{0}}{k^{3}}\int_{0}^{\infty}dt~e^{i\omega t}\left[\sin(k(R_{0}+vt))-k(R_{0}+vt)\cos(k(R_{0}+vt))\right]~.
\end{equation}
This object is formally divergent, but its \emph{on-shell} part, $\omega=\sqrt{k^{2}+m^{2}}$, is \emph{finite}: the divergence takes the form $\delta(\omega-kv)$, and $\omega>kv$ for a massive particle. The energy spectrum of on-shell particles is proportional to the modulus-squared of the FT, and upon removing the off-shell part
($\gamma = 1/\sqrt{1-v^2}$ is the wall Lorentz factor)
\begin{equation}
\begin{split}
   \dfrac{d\langle E\rangle }{dk} = \dfrac{4\pi\gamma^{8}\vert J_{0}\vert^{2}}{k^{4}(k^2+ \gamma^{2}m^2)^4}\Bigg[k^4 \Big(2 k v^3 \cos(k R_0) - R_0 v ((1-v^2) k^2+ m^2) \sin(k R_0)\Big)^2\\+(k^2 + m^2) \Big(((1-3v^2)k^2 + m^2) \sin(k R_0) -k R_0 ((1-v^2) k^2+ m^2) \cos(k R_0)\Big)^2\Bigg]~.
    \end{split}
    \label{eqn:j_mod_squared}
\end{equation}
The qualitative behaviour of this spectrum is shown in figure \ref{fig:schematic_version}: a fast build-up to a first peak (normally, but not always, the global peak), followed by successive peaks of decreasing amplitude. As we discuss in section \ref{sec:finite_wall_thickness} the high-momentum behaviour of this spectrum is an artefact of the approximation of an infinitely thin wall and is not to be trusted, with finite-thickness effects exponentially damping the spectrum above a scale $k_{\rm max} \sim \gamma/R_0$ (this assumes a thick-wall bubble at nucleation; see section \ref{sec:finite_wall_thickness} for details). The position of the global peak in the various kinematic regimes is shown in figure \ref{fig:peak_scaling_for_constant_vel} and further summarized in table \ref{tab:constant_vel_scaling_table}.
Integrating over $k$, the total energy radiated is finite
\begin{equation}
    \begin{split}
       \langle E\rangle  =\dfrac{\pi \gamma^{3}J_{0}^{2}}{6m^{5}}\Bigg[3 e^{-2 \gamma mR_{0}} \left\{\dfrac{7}{\gamma^{6}}-\dfrac{9}{ \gamma ^4}-\left(4 -\dfrac{2}{\gamma^{2}}\right) \left(\dfrac{mR_{0}}{\gamma}\right)^{2}-\left(10 -\dfrac{6}{\gamma^{2}}\right) \dfrac{mR_{0}}{\gamma^{3}}\right\}+\\
       \left(3-\dfrac{9}{\gamma^{2}} +\dfrac{36}{\gamma^{4}}-\dfrac{24}{\gamma^{6}}\right)+4\left(\dfrac{mR_{0}}{\gamma}\right)^{3}+\left(6-\dfrac{12}{\gamma^{2}}\right) \left(\dfrac{mR_{0}}{\gamma}\right)^{2}-\left(24-\dfrac{24}{\gamma^{2}}\right) \dfrac{mR_{0}}{\gamma^{3}}\Bigg]~.
    \end{split}
\label{eqn:energy_theta_constant_vel}
\end{equation}
The two limits $mR_{0}\ll \gamma$ and $mR_{0}\gg \gamma$ have qualitatively different behaviours
\begin{equation} \label{eq:Econstantv_regimes}
    \langle E\rangle \approx \begin{cases}
      \dfrac{\pi v^6 \gamma^3 J_{0}^{2}}{2m^5}~,~~~mR_{0}\ll \gamma   \\
        \\
        \dfrac{2\pi J_{0}^{2} R_{0}^{3}}{3m^{2}}~,~~~ mR_{0}\gg \gamma
    \end{cases}~~.
\end{equation}
(In table \ref{tab:constant_vel_scaling_table} we summarize the scaling of the most important quantities in these two regimes.)  For $mR_0\gg\gamma$, corresponding to case (b) in table \ref{tab:constant_vel_scaling_table}, the energy is proportional to the initial volume, and it is independent of $\gamma$ at leading order even when the bubble wall is moving with large velocity. The reason is that for $mR_0 \gg \gamma$ most of the production is associated with the \emph{appearance} of the bubble at the moment $t=0$, and not with expansion. In other words, it is just the nucleation contribution calculated in the previous section.  

\begin{table}[t!]
    \centering
    \begin{tabular}{|c|c|c|c|}
    \hline
       Cases & $k_{\rm peak}$& $\langle E\rangle$  & $\langle N\rangle$\\
        \hline
     &&&\\
     (a) $mR_{0}\ll \gamma$ & $\sim \frac{\gamma m}{\sqrt{3}}$& $ \frac{v^{6}\gamma^{3} J_{0}^{2}}{m^5}$&$\frac{v^{6}\gamma^{2} J_{0}^{2}}{m^6}$\\
     &&&\\
     \hline
     &&&\\
    (b) $mR_{0}\gg \gamma $& $\sim \frac{1}{R_{0}}$&$\frac{R_{0}^{3} J_{0}^{2}}{m^2}$&$ \frac{R_{0}^{3} J_{0}^{2}}{m^3}$\\
    &&&\\
     \hline
    \end{tabular}
    \caption{Basic characteristics of the spectrum of spin-0 particles of mass $m$ produced by a bubble expanding with constant radial wall Lorentz factor $\gamma$ and initial radius $R_{0}$, in different regimes of $m R_0$: Position of the global peak in the $dE/dk$ spectrum, $k_{\rm peak}$; total radiated energy $\langle E\rangle$; and total number of particles produced, $\langle N\rangle$. The source strength is $J_{0}$. Regime (b), where the particle Compton wavelength is small compared to the Lorentz-contracted initial bubble radius, has almost all particle production occurring at bubble nucleation (see section \ref{sec:bubble_nucleation}). For case (a), the low mass regime, particle production during the expansion is highly efficient and dominates. See also figure \ref{fig:constant_vel_time_regulated}.}
    \label{tab:constant_vel_scaling_table}
\end{table}

On the other hand, when $mR_0\ll\gamma$, i.e. case (a) in  table \ref{tab:constant_vel_scaling_table}, the scaling is much more interesting. Most importantly, the energy scales as the fifth power of the particle Compton wavelength, $1/m^5$, indicating coherent production scaling as the square of the bubble area until production effectively ceases. (This shut-off occurs at bubble radius $\sim \gamma/m$, as we show below.)  The $v^6$ velocity dependence, necessarily present because of the shut-off of production in the adiabatic limit, $v\rightarrow 0$, will play a role in section \ref{sec:back-reaction} when we discuss particle production back-reaction on the bubble wall.

\begin{figure}
\centering
\includegraphics[width=0.7\linewidth]{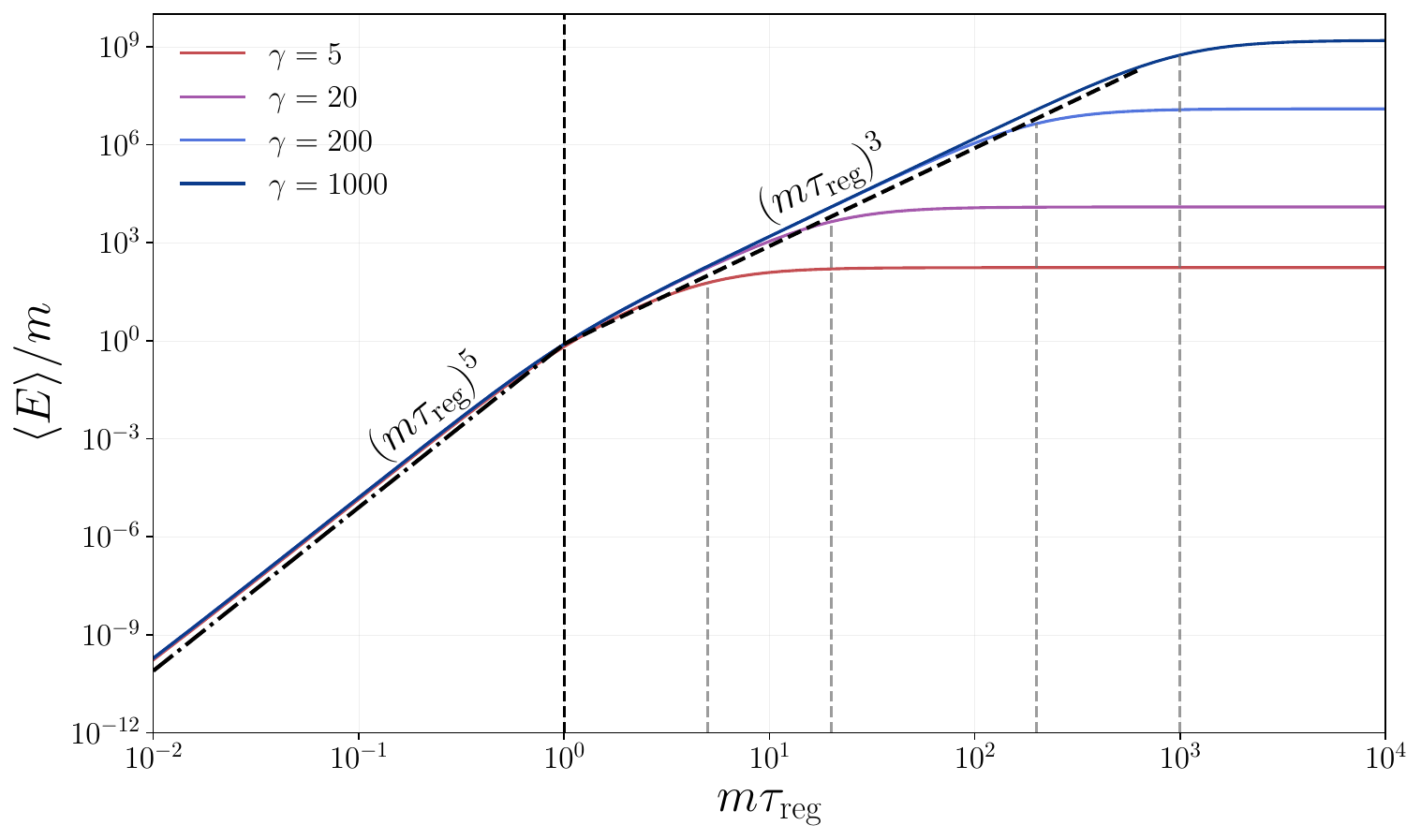}    
    \caption{Total energy $\langle E\rangle$ as a function of time $\taureg$ and $\gamma$ on a log-log scale. The vertical black dashed line shows $m\taureg=1$, while the grey dashed lines correspond to $m\taureg=\gamma$ for the corresponding value of $\gamma$. We take $mR_{0}=0.01$ so we are in the light particle mass regime and $mR_{0}\ll \gamma$ is always satisfied. The figure shows the asymptotic behaviour of the total energy produced in on-shell particles. Initially, the scaling is $\taureg^{5}$. After $\taureg\sim m^{-1}$, the logarithmic slope changes from five to three, eventually going to a $\taureg$-independent quantity.}
\label{fig:constant_vel_time_regulated}
\end{figure}

The behaviour in the light-mass regime, $mR_0\ll\gamma$, is further elucidated by the numerical results of figure \ref{fig:constant_vel_time_regulated}. Here we show the energy produced as a function of time $\taureg$, extracted using an adiabatic regulator of the form $\exp(-t/\taureg)$. The black dashed line marks $\taureg=m^{-1}$, and the grey vertical lines mark $\taureg=\gamma m^{-1}$ for each value of $\gamma$. We first observe that each curve flattens at $\taureg\sim\gamma m^{-1}$, up to an $\mathcal{O}(1)$ factor. This indicates that at time $\gamma m^{-1}$ production becomes very inefficient and effectively shuts off, explaining the resulting finite total energy. The plot also shows the derivative of $\log \langle E\rangle$ with respect to $\log \taureg$. At early times, $\taureg < m^{-1}$, when the mass is effectively irrelevant, the total energy grows as $\taureg^{5}$. This corresponds to a coherent production rate that scales as the \emph{square of the bubble wall area}. Between $m^{-1}$ and $\gamma m^{-1}$, the scaling decreases to $\taureg^{3}$, indicating a shift from coherent production to incoherent production proportional to the wall area. Finally, after $\taureg\sim \gamma m^{-1}$, the production effectively shuts off.

\subsection{Physics of production cessation}
\label{subsec:cessation}
Equation \eqref{eqn:energy_theta_constant_vel} shows that even an eternally expanding constant-velocity wall deposits only a finite amount of energy into particle production. This finiteness carries two lessons. First, production cannot be a purely local process --- otherwise one could pass to the local rest frame of the wall, where the source is static, and conclude that no particles are produced. Second, production must become inefficient after some characteristic time. Figure \ref{fig:constant_vel_time_regulated} shows numerically that this occurs at bubble radius $R\sim\gamma/m$; we now explain both points.

At each point of the wall, for the constant velocity case, a local inertial rest frame exists, but there is no single inertial frame in which an entire ``radiation formation region'' of size $\sim 1/m$ is static when the wall curvature is appreciable
\begin{equation}
    R_{\rm rest}\lesssim \dfrac{1}{m}~~.
\end{equation}
Here $R_{\rm rest}$ is the radius of curvature observed in the local rest frame. To see how this translates into a constraint on the cosmological ``lab'' frame bubble radius, consider how the curvature changes as we transform to the local frame (which, without loss of generality, we take to be that of the north pole of the bubble). In the lab frame, in which the bubble wall expands with constant velocity $v$, the wall satisfies
\begin{equation}
    x^{2}+y^{2}+z^{2}=(R_{0}+vt)^{2}~~.
\end{equation}
In the local frame of the north pole, with primed coordinates, this becomes
\begin{equation*} x'^{2}+y'^{2}+\left(\gamma\left(z'+vt'\right)+R_{0}\right)^{2}=\left(R_0+v\gamma(t'+vz')\right)^{2}~~,
\end{equation*}
which can be rearranged as
\begin{equation*}
    x'^{2}+y'^{2}+(1+v^{2})\left(z'+\dfrac{1}{1+v^{2}}\left(vt'+\dfrac{R_{0}}{\gamma}\right)\right)^{2}=\dfrac{1}{(1+v^{2})}\left(vt'+\dfrac{R_{0}}{\gamma}\right)^{2}~~.
\end{equation*}
Using the standard mean curvature formula \cite{Mathsformula}, we find that the radius of curvature at the north pole is
\begin{equation}
    R_{\rm rest}=\left(vt'+\dfrac{R_{0}}{\gamma}\right)=\dfrac{R_{\rm lab}}{\gamma}~~,
\end{equation}
where, in the last equality, we have used that for the observer at the north pole $t'= t/\gamma$. Hence particle production becomes \emph{inefficient} once
\begin{equation}
    R_{\rm lab}\gtrsim  \dfrac{\gamma}{m}~~,
\label{eqn:flat_wall_condition}
\end{equation}
which agrees with our analytic and numerical results.

\subsection{Expansion at constant proper acceleration}
\label{subsec:constant_proper_acceleration}

We now consider bubbles expanding with constant proper acceleration up to a time $\tauacc$. Two types of trajectory include such an accelerated phase: (a) runaway, in which the bubbles accelerate until they collide; and (b) acceleration followed by a constant-velocity phase, as a consequence of friction. In either case, at the end of the FOPT the bubbles collide and the PT completes after complicated collision dynamics. As collision is a multi-bubble process, a source profile for a single bubble cannot capture it. Further, depending on the two scenarios, $\tauacc$ can be either the time of phase transition completion $\betagw^{-1}$ in the case of runaway or simply the switch-over time from accelerated phase to the constant velocity phase. Below we discuss the production of particles captured by a single bubble profile in these two cases.

\subsubsection*{Runaway Scenario}
\begin{figure}
\centering
\begin{subfigure}{0.46\linewidth}
 \centering
\includegraphics[width=\linewidth]{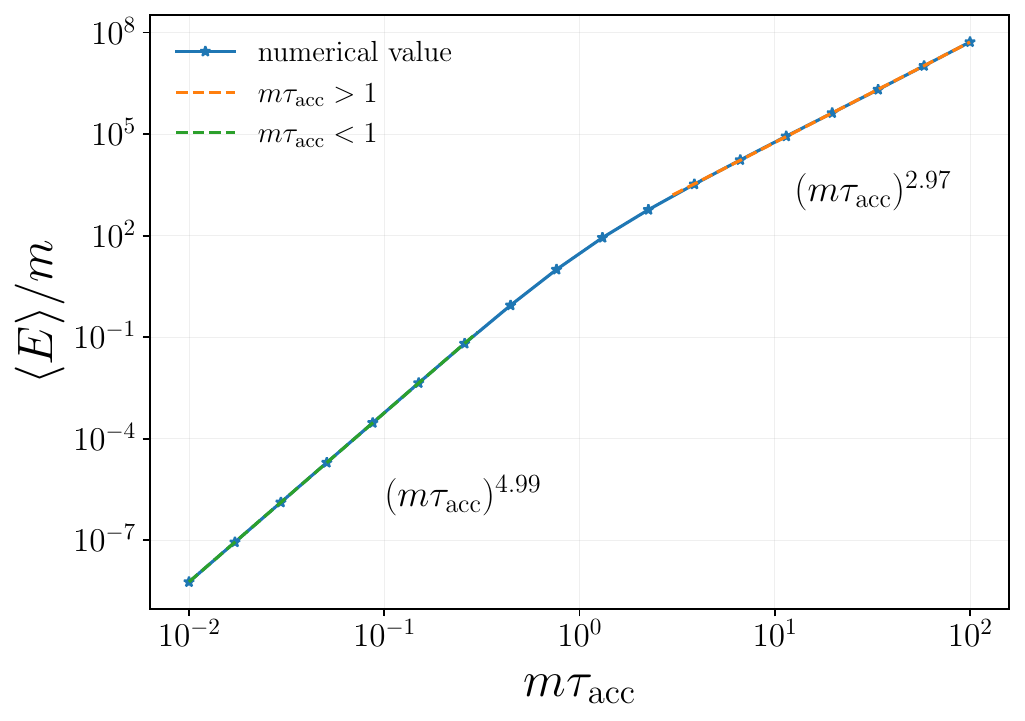}   
\end{subfigure}
  \begin{subfigure}{0.475\linewidth}
 \centering \includegraphics[width=\linewidth]{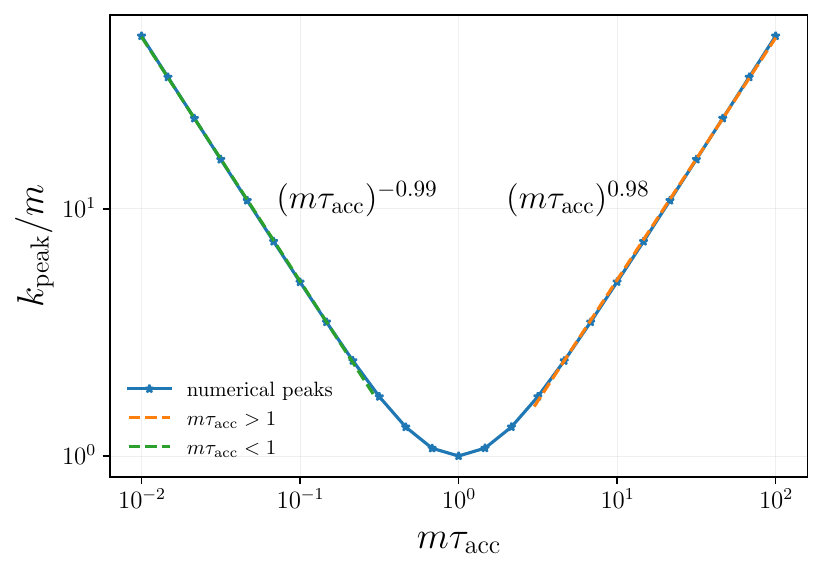}   
\end{subfigure}  
    \caption{\textbf{Left}: Energy scaling in the runaway case as a function of FOPT completion time $\tauacc$. \textbf{Right}: Scaling of the distribution peak position. Numerical results agree with the asymptotic scaling laws of table \ref{tab:runaway_exponential_regulator_scaling}. Production rate is coherently enhanced for $m \tauacc < 1$, and incoherent at later times. Peak position shows a switch-over of behaviour at $m \tauacc\sim 1$.}
\label{fig:exponential_regulated_runaway}
\end{figure}

In this scenario the bubbles accelerate until they collide, completing the transition. A single-bubble source $J(t,r)$ that eventually turns off therefore captures the expansion phase but also part of the collision. We emphasize again that collision is a multi-bubble process, so this profile cannot capture the full collision dynamics; at the same time, the expansion and collision phases cannot be cleanly separated. For a runaway scenario, a possible function that models the source and captures the key aspects of expansion and collision is 
\begin{equation}
J(t,r)=J_{0}\Theta(t)\Theta\left(\sqrt{R_{0}^{2}+t^{2}}-r\right)\dfrac{\left(1-\tanh((t-\tauacc)/\widetilde{\tau})\right)}{1+\tanh(\tauacc/\widetilde{\tau})}~~.
\label{eqn:tanh_regulated}
\end{equation}
In the profile above, $\tauacc$ denotes the time at which expansion ends and the bubbles collide, while $\widetilde{\tau}$ sets the duration over which the transition completes. Intuitively, this is an envelope-type approximation to the completion of the transition, agnostic to high-momentum (ultraviolet) details. In the limit $\widetilde{\tau}\rightarrow 0$ the regulator reduces to a Heaviside function, whereas for $\widetilde{\tau}\sim \tauacc$ it behaves exponentially. Generically, then, two scales determine the features of the spectrum. The hierarchy between $\widetilde{\tau}$ and $\tauacc$ is model-dependent, set in part by the collision dynamics.
\begin{figure}[t!]
    \centering
    \includegraphics[width=0.7\linewidth]{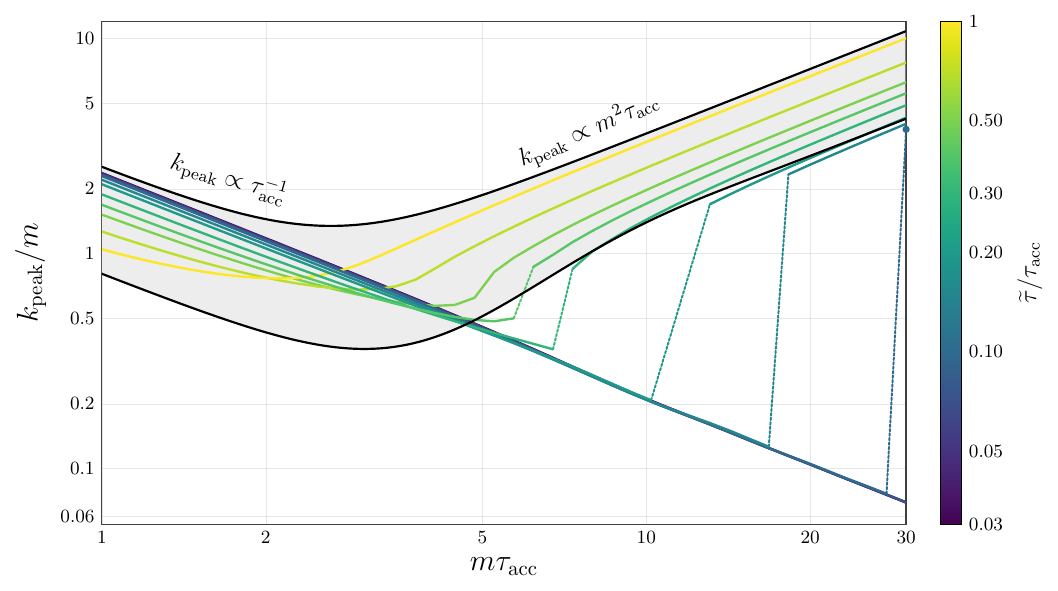}
    \caption{Behaviour of the position of the spectrum peak, $k_{\rm peak}$, for the source  equation~\eqref{eqn:tanh_regulated}, as a function of $\tauacc$ for different values of $\widetilde{\tau}/\tauacc$. For small $\widetilde{\tau}$, the peak occurs at low momentum for a wide range of $\tauacc$. For $\widetilde{\tau}\sim \tauacc$, the peak scales as equation \eqref{eqn:k_peak_tanh}.}
    \label{fig:tanh_regulator_peak}
\end{figure}
Figure \ref{fig:tanh_regulator_peak} shows that the peak of the spectrum depends on both the ratio $\widetilde{\tau}/\tauacc$ and $m\tauacc$. Generically, when $\widetilde{\tau}\sim \tauacc$ up to order-one factors, the peak behaves as
\begin{equation}
    \dfrac{k_{\rm peak}}{m}\sim\begin{cases}
        \dfrac{1}{m\tauacc}~~~m\tauacc<m\tau_{\star}~,~~\\
        \\
        m\tauacc~~~m\tauacc>m\tau_{\star}~~,
    \end{cases}
    \label{eqn:k_peak_tanh}
\end{equation}
where $m\tau_{\star}\sim \mathcal{O}(1)$ is the location where the behaviour changes, whose exact value depends on the ratio $\widetilde{\tau}/\tauacc$. Already this simple setup shows that collision dynamics can have an interesting impact on the spectrum. To consider a simple model, we choose $\widetilde{\tau}\sim \tauacc$. This is a natural expectation for systems where a single scale drives the transition. (There are, though, cases where formation of oscillon-like phases can lead to longer completion times that need separate modelling.) In this case the source can formally be approximated as
\begin{equation}
J(t,r)=J_{0}\Theta(t)\Theta\left(\sqrt{R_{0}^{2}+t^{2}}-r\right)e^{-t/\tauacc}~.
\label{eqn:acc_source}
\end{equation}
As before, we will be interested in a scenario in which the bubble accelerates to relativistic velocities, corresponding to $R_{0}\ll \tauacc$. The energy and peak of the spectrum depend on $m$ and $\tauacc$. Since the calculation closely parallels that of section \ref{subsec:constant_velocity_and_the_regulator}, we simply show the dependence of $\langle E\rangle$ and $k_{\rm peak}$ on $m\tauacc$ in figure \ref{fig:exponential_regulated_runaway} and summarize the scaling laws in table~\ref{tab:runaway_exponential_regulator_scaling}. The scaling of $k_{\rm peak}$ shows the same qualitative behaviour as the more general setup with $\widetilde{\tau}\sim \tauacc$ of figure \ref{fig:tanh_regulator_peak}. The rate of production is coherently enhanced for $m \tauacc < 1$, and incoherent at later times, although now, unlike the constant wall velocity case, the production never switches off. The total energy produced scales as the total volume $\propto\tauacc^{3}$, as expected. Evidently, if the acceleration epoch were infinite in duration, we would find a divergent result for particle production. For a cosmological FOPT this divergence is spurious, as it is cut off by the completion of the transition.
\begin{table}
    \centering
    \begin{tabular}{|c|c|c|c|}
    \hline
      Scales  (with $mR_{0}\ll 1$) & $k_{\rm peak}$ & $\langle E\rangle$ &$\langle N\rangle$ \\
      \hline
      & & &\\
       $m\tauacc<1$ &$0.52/\tauacc$  & $1.44 ~J_{0}^{2}\tauacc^{5}$& $1.44~J_{0}^{2}\tauacc^{6}$\\
      & & &\\
      \hline
      & & &\\
      $m\tauacc>1$ & $0.54~m^{2}\tauacc$ &$ 1.54~J_{0}^{2}\tauacc^{3}m^{-2}$& $1.44~ J_{0}^{2}\tauacc^{2}m^{-4}$\\
      & & &\\
      \hline
    \end{tabular}
    \caption{$\tauacc$ dependence of $k_{\rm peak}$, $\langle E\rangle$ and $\langle N\rangle$ for the runaway case, assuming $mR_{0}\ll 1$ and $R_{0} \ll \tauacc$ and the source profile as given in equation \eqref{eqn:acc_source}.}
\label{tab:runaway_exponential_regulator_scaling}
\end{table}
\subsection*{The accelerated patched trajectory with hard cut-off}
\label{sec:hard_cutoff}
Apart from the runaway scenario, it is common for FOPT bubbles to eventually achieve a constant terminal velocity as a consequence of velocity-dependent friction. To understand this scenario in a model-independent way, we consider a very simple case where the accelerated phase terminates and a constant velocity phase starts at time $\tauacc$, with the trajectory of the source and its first derivative in time patched continuously. Although the physical trajectory is $C^{\infty}$, here we consider a $C^{1}$ function. This affects the tail of the spectrum which, as we will see in section \ref{sec:finite_wall_thickness}, is in any case already unreliable due to finite wall-thickness effects. The source in question is
\begin{align}
   & J(t,r)=\Theta(t)[J_{\rm acc}(t,r)+J_{\rm cst-vel}(t,r)]~,~~
    J_{\rm acc}(t,r)=J_0\Theta(\tauacc-t)\Theta\left(\sqrt{R_{0}^{2}+t^{2}}-r\right)\nonumber\\
   &J_{\rm cst-vel}(t,r)= J_0\Theta(t-\tauacc)\Theta\left(R_{\rm acc}+v_{\rm term}(t-\tau_{\rm acc})-r\right)~,~R_{\rm acc}=\sqrt{R_{0}^{2}+\tauacc^{2}}~~.
    \label{eqn:patched_trajectory}
\end{align}
The FT of equation \eqref{eqn:patched_trajectory} is a sum of the individual FTs of the accelerated part and the constant velocity part with a phase difference coming from the Heaviside function. The modulus squared of the total FT is therefore a sum of two modulus squares and an interference term. Integrating then gives the total energy as a sum of the two contributions plus an interference term:
\begin{equation*}
    \langle E\rangle=\langle E\rangle_{\rm acc}+\langle E\rangle_{\rm cst-vel}+2\cos(\theta)\sqrt{\langle E\rangle_{\rm acc}\langle E\rangle_{\rm cst-vel}}~~,
    \label{eqn:separate_energy_definition}
\end{equation*}
where $\theta$ is an interference angle arising from a complicated integral, and the subscripts ``acc'' and ``cst-vel'' refer to the accelerated and constant-velocity contributions. If one of $\langle E\rangle_{\rm acc}$ or $\langle E\rangle_{\rm cst-vel}$ is significantly larger than the other, say, by an order of magnitude, then the larger one will dominate the scaling behaviour of the total energy $\langle E\rangle$. Let us first check whether production continues once the bubble enters the constant-velocity regime. Since we are considering $mR_{0}\ll 1$,
\begin{equation*}
     \sqrt{R_{0}^{2}+\tauacc^{2}}\ll \frac{\gammaterm}{m}=\dfrac{\sqrt{R_{0}^{2}+\tauacc^{2}}}{mR_{0}}~~,
\end{equation*}
so particle production does not immediately cease in the constant velocity phase. Using our expression for $\gammaterm$ and the results in Tables~\ref{tab:constant_vel_scaling_table} and \ref{tab:runaway_exponential_regulator_scaling} we find that the ratio between the two terms is 
\begin{equation*}
    \dfrac{\langle E\rangle_{\rm acc}}{\langle E\rangle_{\rm cst-vel}}\lesssim (mR_{0})^{3}\ll 1~~.
\end{equation*}
Hence, the energy produced in the constant velocity phase of the trajectory is much larger than that produced in the initial acceleration phase. This is intuitive: for both contributions the energy is proportional to a bubble volume: that at the end of acceleration for $\langle E\rangle_{\rm acc}$, and that at the shut-off radius $\gammaterm/m$ for $\langle E\rangle_{\rm cst-vel}$. Since the latter greatly exceeds $\tauacc$, $\langle E\rangle_{\rm cst-vel}$ dominates.
 
\subsubsection*{The scaling}
Consider now the scaling of the spectrum and total energy for a FOPT completing in a time $\betagw^{-1}$, in two scenarios: runaway, and terminal velocity with production shutting off before collision (corresponding to $\gammaterm=m/\betagw$). From the earlier discussions and Figures \ref{fig:constant_vel_time_regulated} and \ref{fig:exponential_regulated_runaway} we have learned that the energy in the particles scales as the volume. Thus, whether in the runaway scenario, where the bubble expands up to radius $\betagw^{-1}$, or in a terminal-velocity scenario in which the would-be shut-off radius $\gammaterm/m$ exceeds $\betagw^{-1}$, production during expansion is set by the final radius $\betagw^{-1}$. In the case $mR_0\ll1$, we can therefore summarize the result for a bubble that accelerates for a time $\tauacc$, reaching $\gammaterm=\tauacc/R_0$, and collides at a time $\betagw^{-1}$, by the following scaling in energy
\begin{equation}
    \langle E\rangle \approx \dfrac{\pi}{2}\dfrac{J_{0}^{2}\gammaeff^{3}}{m^{5}}~,~~~\gammaeff=\text{min}\left\{\gammaterm,\dfrac{m}{\betagw}\right\}~~.
    \label{eqn:final_energy_scaling}
\end{equation}

Besides the total energy, we are also interested in the spectrum of the particles produced. For the runaway scenario modelled by equation \eqref{eqn:acc_source}, the peak position goes as $k_{\rm peak}=m^{2}\betagw^{-1}$, while for terminal velocity with $\gammaterm\leq m/\betagw$ it scales as $k_{\rm peak}=\gammaterm m$. In the limiting case the two scalings coincide:
\begin{equation}
    k_{\rm peak}\approx 0.5\, m\gammaeff~,~~~\gammaeff=\text{min}\left\{\gammaterm,\dfrac{m}{\betagw}\right\}~~.
\end{equation}
We remind the reader that these results hold for the class of regulators with $\widetilde{\tau}\sim \tauacc$; markedly faster or slower completion following collision requires separate study.

Notice that the above scaling relies on the bubbles reaching a maximum radius $\betagw^{-1}$, which leads to an effect that may at first seem surprising. If production has not ceased by the time the bubbles collide, then $k_{\rm peak}/m\sim m/\betagw$: for a fixed transition duration, the way to obtain a \emph{predominantly} relativistic spectrum, i.e.~dark radiation, is to increase the particle mass. The caveat is that predominant does not mean substantial: although a heavier particle yields a larger relativistic fraction, the total number of particles produced is suppressed by a factor of $m^{-3}$. A substantial amount of dark radiation in the runaway case therefore requires a very large maximum radius $\betagw^{-1}$, which increases both the typical momentum and the total energy produced.

\subsection{Finite wall thickness effects}
\label{sec:finite_wall_thickness}

So far we have modelled the bubble wall profile as a Heaviside $\Theta$-function. Physical walls, however, have finite thickness, which modifies the spectrum of produced particles at momenta above the inverse of Lorentz-contracted wall thickness in the FRW frame.

The large-$|k|$ fall-off of the spectrum of produced particles is set by the decay of the Fourier transform squared. For functions with $n$ integrable derivatives the FT has an asymptotic $1/|k|^n$ fall-off, reflected in the slow polynomial decay of the spectra at large $|k|$. On the other hand, for physical walls we expect the profile to be analytic, which gives a fall-off faster than any power of $|k|$.  If we model the wall profile by $f(r)= \tanh(r/w)$, where $w$ parametrises the $\gamma$-dependent wall thickness, then the integral equation \eqref{eqn:spherically_sym_fourier}
\begin{equation}
    {\tilde f}(k)=4\pi \int_0^\infty dr r^2 f(r) \frac{\sin(kr)}{kr}
\end{equation}
has, after the subtraction of the off-shell constant asymptote, as its large-$|k|$ behaviour
\begin{equation}
    {\tilde f}_{\rm reg}(k) \sim \frac{\rm const}{k} \exp(-\pi wk/2)
\end{equation}
determined by the position $r=\pm i \pi w/2$ of the nearest poles of $\tanh(r/w)$ in the complex plane. Thus, we expect for general smooth wall profiles that the $k$-spectra of the 
produced on-shell particles will be exponentially cut off beyond $k_{\rm max}\simeq 1/w$, where $w$ is the appropriate wall thickness. What value of $w$ should we take? A conservative estimate follows from assuming that the bubble at nucleation is thick-walled with $w_{\rm initial} \simeq R_0$. This is then contracted by the appropriate Lorentz factor. For the constant velocity trajectory case we therefore simply impose 
\begin{equation}
    k<k_{\rm max} = \frac{\gamma}{R_0}~.
\end{equation}
For the accelerating bubble wall, we must choose the $\gamma$ appropriate for the physical question. For our discussion of dark radiation to hold, we require that the Lorentz factor $\gamma_{\rm peak}$ producing the dominant peak of the momentum distribution --- which in turn determines, to good approximation, the amount of ultra-relativistic radiation --- satisfies
\begin{equation}
    \gamma_{\rm peak} > k_{\rm peak} R_{0}~~.
\end{equation}
Here, again, we make the conservative assumption that the initial bubble is thick-walled.
here
\section{Production of axion-like particles during confining phase transitions}
\label{sec:alp_pheno_case}

The particle production mechanism of section \ref{sec:Particle production from expanding bubbles} finds a natural realization in ALPs coupled to a hidden-sector Yang-Mills theory undergoing a deconfinement-confinement FOPT: as we show below, the physical ALP excitation couples linearly to a source that changes value across the bubble wall --- precisely the setup of equation \eqref{eqn:general_lagrangian}. Both ingredients are independently well motivated: ALPs are a generic prediction of UV completions such as string theory \cite{Svrcek:2006yi, Arvanitaki:2009fg}, while hidden confining sectors are ubiquitous in extensions of the Standard Model, their first-order transitions being an active target of gravitational wave searches. As will soon be apparent, however, the ALP cannot be the QCD axion. For definiteness we consider a pure $SU(\Nc)$ theory without matter fields; for $\Nc\geq3$ this theory undergoes a first-order deconfinement-confinement thermal PT. (A brief discussion of $SU(\Nc)$ FOPT dynamics is in appendix \ref{appendix:transition_dynamics}.)

\subsection{ALP coupling, potential, and production}
Generically, this sector will also be endowed with a bare $\theta$-term
\begin{equation}
    {\mathcal L}_{\theta} = \theta_0\frac{g^2}{32\pi^2} G \widetilde{G}~~.
\end{equation}
In the presence of an uncancelled $\theta$-term the confining regime has both CP-even \emph{and} CP-odd VEVs\footnote{Unlike the $\mathbb{Z}_{\Nc}$-centre-symmetry sensitive Polyakov-loop, these are \emph{not} strictly defined order parameters for the FOPT, as they have a non-zero, though different, value in the deconfined phase.}
\begin{equation}
    g^{2} \Nc \langle G G\rangle \sim \Lambda^{4}~,\qquad g^{2} \Nc \langle G \widetilde{G}\rangle \sim f(\theta_{0})\Lambda^4~,
\end{equation}
where $f(\theta_0)$ is an ${\cal O}(1)$ odd periodic function satisfying $f(0)=0$, and $\Lambda$ is the Yang-Mills dynamical scale. The dynamics of the PT at $\theta_0\neq 0$ is far from fully understood; however, studies indicate that the strength of the deconfinement-confinement FOPT \emph{increases}, at least for $\theta_0$ small \cite{Bonati:2013tt}. Moreover, across the bubble wall that separates the exterior deconfined phase from the interior confined phase, the values of both $\langle G G\rangle$ \emph{and} $\langle G \widetilde{G}\rangle$ jump (for $\theta_0\neq 2\pi n$).  These statements are supported by analytic studies of the confining regime in controlled setups, such as supersymmetric gauge theories~\cite{Veneziano:1982ah} or Yang-Mills theories defined on $\mathbb{R}^{3}\times S^{1}$~\cite{Aitken:2018mbb}. 

From a fundamental UV perspective, e.g.~in string constructions, an axion-like field, $a(x)$, is expected to exist in this sector, coupling in the usual way (we set the domain wall number $N_{\rm DW}=1$ for simplicity)
\begin{equation}
    {\mathcal L}_{aG{\tilde G}} = \dfrac{g^{2}}{32\pi^2}\frac{a}{f_{a}} G \widetilde{G}~~.
\end{equation}
Were this the whole story, this interaction would give rise to a non-perturbatively generated potential $V_{\rm np}(a)$, the physical axion excitation would acquire a mass $m_a\sim\Lambda^2/f_a$, and $\theta_0$ would be \emph{cancelled} by the usual Peccei-Quinn (PQ) mechanism~\cite{Peccei:1977hh, Peccei:1977ur, Weinberg:1977ma, Wilczek:1977pj} (at least in the low-temperature regime $H^2 \sim T^4/M_{\rm pl}^2 \ll m_a^2$, $H$ being the Hubble scale). In that case there would be no large confined-phase bubble source $\langle G \widetilde{G}\rangle$ coupled to the physical axion excitation, and likely also a significant suppression of the coupling in the deconfined phase.

Fortunately, however, it is also to be expected that there exist additional contributions that break the continuous PQ shift symmetry, some arising from far UV effects, possibly quantum gravity or string theory. This is the analogue, transcribed to this sector, of the well-known axion quality problem of the QCD axion \cite{Dine:1986bg, Kamionkowski:1992mf, Barr:1992qq, Holman:1992us}. In the QCD axion case one usually has to work to ensure that the axion is adequately protected from these effects, so that the effective $\theta$-parameter remains sufficiently small. \footnote{For instance, this can be achieved by employing accidental protection arising from discrete gauge symmetries \cite{Kamionkowski:1992ax, Babu:2002ic, Lee:2011dya, Harigaya:2013vja, Duerr:2017amf, Bhattiprolu:2021rrj}, from locality in extra-dimensional constructions \cite{Choi:2003wr, Svrcek:2006yi, Arvanitaki:2009fg,Petrossian-Byrne:2025mto} or having axions as composite objects \cite{Kim:1984pt,Chun:1991xm,Randall:1992ut,Flacke:2006ad,Redi:2016esr, Lillard:2017cwx, Gavela:2018paw, Cox:2019rro, Ardu:2020qmo, Yin:2020dfn, Contino:2021ayn, Cox:2023dou, Nakagawa:2024kcb, Gherghetta:2025fip, Gherghetta:2025kff, Agrawal:2025mke}.} In contrast, here we want $\theta_0$ to be imperfectly cancelled, leaving a residual ${\cal O}(1)$ CP-violating angle. Indeed, it would be quite surprising if \emph{all} axion-like particles were protected from UV shift-symmetry breaking effects.  So, in the far IR, we assume the potential is
\begin{equation}
    V(a)=V_{\rm np}+\vqp~~.
\end{equation}
Here the ``quality problem potential'' $\vqp$ is taken to be simply
\begin{equation}
    \vqp=-\muqp^{4}\cos\left(\dfrac{a}{f_{a}}-\delta\right)~~.
\end{equation}
where $\muqp$ is the ``quality problem scale'' and $\delta\sim{\mathcal O}(1)$ is an unknown UV angle. We take $\muqp>\Lambda$, so that $\vqp$ dominantly sets the axion VEV, approximately equal to $f_a\,\delta$. \footnote{One can also consider the case $\muqp<\Lambda$ where the residual CP-violating $\theta$-angle, $\theta_{\rm eff}$, is suppressed. Phenomenologically interesting scenarios exist in this regime as long as $\theta_{\rm eff}\not\ll 1$.} In the confining regime, and at low momentum, the effective Lagrangian describing this system is built from gauge-invariant objects associated with the glueballs, which are the appropriate degrees of freedom below $\Lambda$. One cannot exactly calculate this effective Lagrangian, but one can use a spurion analysis to write down its general form, including $a(x)$, at scales below $\Lambda$, in the spirit of the QCD chiral Lagrangian. (We direct readers to \cite{Carenza:2024avj,Carenza:2024qaq, Rosenzweig:1979ay, Schechter:1980ak} for more details.) In particular, leading contributions arise from the $0^{++}$ glueball interpolating operator $\mathcal{H}$ given by $\mathcal{H}^{4}\equiv -(\beta(g)/2g) \Tr(G^{2})$, and the pseudoscalar $0^{-+}$ glueball interpolating field $\mathcal{A}$ defined via $\mathcal{A}\mathcal{H}^{3}\equiv \Tr(G\widetilde{G})$ \cite{Carenza:2024avj, Carenza:2024qaq}. One has
\begin{equation}
    \mathcal{L}_{\rm eff}=\frac12 (\partial\mathcal{H})^{2}+\frac12 (\partial\mathcal{A})^{2}-V_{\rm eff}(\mathcal{H},\mathcal{A})+\ldots~,
\end{equation}
and, crucially, when the effective $\theta$-angle $\theta_{\rm eff}\neq 0$, a term $\theta_{\rm eff}\mathcal{A}\mathcal{H}^{3}\in V_{\rm eff}(\mathcal{H},\mathcal{A})$, which induces the coupling of the physical axion via the replacement $\theta_{\rm eff}\rightarrow \theta_{\rm eff}+a/f_{a}$. The potential $\vqp$ sets the VEV for the axion, which in turn sets the VEV for $G\widetilde{G}$ in the confined phase. Thus, including the effective CP-odd term in the effective glueball Lagrangian, we have
\begin{equation}
    \mathcal{L}_{\rm eff}\supset \dfrac{a}{f_{a}}\Lambda^{4} f(\theta_{\rm eff})~~,
    \label{eqn:effective_linear_term}
\end{equation}
where now $a$ is the axion field around the VEV and $\theta_{\rm eff}\simeq\theta_{0}+\delta$ is the effective $\theta$-parameter, assumed ${\cal O}(1)$ (for $\muqp>\Lambda$). Thus, \emph{the physical axion quanta are linearly coupled to a source that changes value across the deconfined-confined bubble wall.} This implies
\begin{equation}
    J_{0}\sim \dfrac{\Lambda^{4}}{f_{a}},~~~~m_{a}\sim \dfrac{\muqp^{2}}{f_{a}}~~.
    \label{eqn:axion_parameters}
\end{equation}
After the transition, we require the axion to be light, $m_{a}<\Lambda$, giving the constraint
\begin{equation}
    1<\dfrac{\muqp}{\Lambda}<\sqrt{\dfrac{f_{a}}{\Lambda}}~~.
\label{eqn:prelim_condition_on_pq_breaking_scale}
\end{equation}
To check the weak-coupling condition, we expand the effective Lagrangian around $\theta_{\rm eff}$ and find that the $n$th-order ALP self-coupling scales as
\begin{equation}
   \lambda_n\lesssim\frac{\muqp^4}{f_a^n}~~.
\end{equation}
Imposing equation \eqref{eq:weak_source} gives the weak-source condition
\begin{equation*}
    2\frac{\lambda_n}{n!}\frac{J_0^{n-2}}{m_a^{2n-2}}\lesssim\frac{2}{n!}\left(\frac{\Lambda}{\muqp}\right)^{4(n-2)} \ll  1~~,
\end{equation*}
which is automatically satisfied for all $n\geq3$ when $\muqp>\Lambda$. By equation \eqref{eq:mass_correction}, the mass corrections are then bounded by
\begin{equation*}
    \frac{\Delta m^2_a}{m^2_a}\leq \left(\frac{\Lambda}{\muqp}\right)^4~~,
\end{equation*}
so $m_a$ can be identified with the physical mass of the produced particle.
\begin{table}
    \centering
    \begin{tabular}{|c|c|c|}
    \hline
        Quantity & Symbol in  section \ref{sec:Particle production from expanding bubbles}  & Parameter in the model  \\
        \hline
       Mass& $m$  & $m_{a}\approx \muqp^{2}/f_{a}$\\
       \hline
       Source strength& $J_{0}$ & $\Lambda^{4}/f_{a}$\\
       \hline
       Initial size& $R_{0}$ & $\Lambda^{-1}$\\
       \hline
       Runaway time& $\tauacc$ & $\betagw^{-1}$\\
       \hline
       Self coupling& $\lambda_n$ & $\lambda_n\lesssim\muqp^4/f_{a}^n$\\
       \hline
       \hline
       \multicolumn{3}{|c|}{Choice of Parameters} \\
       \hline
       \multicolumn{3}{|c|}{$f_{a}=\{10^{10},10^{16}\}\text{GeV}~,~~\betagw/H_{\rm n}=10^{4}$~,~~$\betagw<m_{a}<\Lambda$} \\
       \hline
    \end{tabular}
    \caption{Mapping between the quantities in section \ref{sec:Particle production from expanding bubbles} and the parameters of the model.}
    \label{tab:mapping_of_parameters}
\end{table}

Turning to ALP production, we take the acceleration time in the runaway case to be $\betagw^{-1}$, the PT completion time. As we saw in section \ref{subsec:constant_proper_acceleration}, for $m_{a}/\betagw<1$ we have coherent production of particles until collision. If the walls collide while production is still coherent, collision dynamics could significantly affect production throughout the bubble volume; to be conservative, we therefore restrict to $m_{a}/\betagw>1$. Using the definition of $m_{a}$ in equation \eqref{eqn:axion_parameters} and a choice of $\betagw$, this condition, together with equation \eqref{eqn:prelim_condition_on_pq_breaking_scale}, can be expressed as
\begin{equation}
     \max\left\{1,\left(\dfrac{\betagw}{H_{\rm n}}\dfrac{f_{a}}{M_{\rm pl}}\right)^{1/2}\right\}<\dfrac{\muqp}{\Lambda}<\sqrt{\dfrac{f_{a}}{\Lambda}}~~.
     \label{eqn:parameter_space_for_mu_lambda}
\end{equation}
Our numerical analysis in section \ref{subsec:spectrum_analysis} uses benchmark values, $f_{a}=\{10^{10},10^{16}\}~\text{GeV}$, and $\betagw/H_{\rm n}=10^{4}$. For the map between the model parameters and the quantities of section~\ref{sec:Particle production from expanding bubbles}, see table \ref{tab:mapping_of_parameters}.

\subsection{Comparison with the population produced by freeze-in}
\label{appendix:reheat}
Because the axion couples to the gauge sector undergoing the transition, a population of axions is inevitably generated before the PT via freeze-in. Before studying the spectrum produced by bubble expansion, we verify that this pre-existing population is subdominant: we show that for a wide range of reheating temperatures $T_{\rm reheat}$, the freeze-in abundance is negligible compared to that produced during the transition.

After the PT, axion production via scattering of glueballs will be heavily suppressed because the glueballs are heavy, and hence the dominant thermal population should be generated in the deconfined phase. \footnote{Note that post-PT the metastable glueballs can decay into the lightest CP-even and CP-odd states which then can decay to axions. This extra potential complication is in fact quite model-dependent, so we do not consider it here, briefly returning to the possibility in section \ref{sec:discussion}.} Requiring that the axions produced during the PT dominate over freeze-in places an upper bound on the reheating temperature.

Suppose that at the reheating temperature $T_{\rm reheat}$ the axion abundance vanishes while the gauge sector is thermalised. At leading order, the contribution will come from a process of the form
\begin{equation*}
    \text{gluon}+\text{gluon}\xrightleftharpoons[]{} \text{gluon}+\text{axion}~~,
\end{equation*}
and the thermally averaged Boltzmann equation takes the form
\begin{equation}
    \dfrac{1}{a^{3}}\dfrac{d (n_{\rm alp}a^{3})}{dt}=\Gamma_{gg\rightarrow ga}(T)\left(1-\dfrac{n_{\rm alp}}{n_{\rm alp}^{\rm eq}}\right)~~,
    \label{eqn:boltzmann_alp}
\end{equation}
where $\Gamma_{gg\rightarrow ga}(T)$ is the reaction density of the scattering above. On the basis of scaling arguments, the reaction density takes the form
\begin{equation}
    \Gamma_{gg\rightarrow ga}(T)\sim\dfrac{(\alpha_{s}\Nc)^{3}}{f_{a}^{2}}T^{6}~~.
\end{equation}
Further, as we want the scale hierarchy $m_{a}<\Lambda<T_{\rm reheat}$, the axions produced thermally in the pre-transition epoch behave as radiation, so $n_{\rm alp}^{\rm eq}\sim T^{3}$. For the freeze-in calculation to be valid throughout, the axions must not thermalise, i.e., $n_{\rm alp}(T)/n_{\rm alp}^{\rm eq}(T)\ll 1$. This is satisfied when 
\begin{equation*}
    \dfrac{\Gamma_{gg\rightarrow ga}(T)}{n_{\rm alp}^{\rm eq}(T)}\ll H(T)~~.
\end{equation*}
In a radiation-dominated era, $H\sim T^{2}/M_{\rm pl}$, and requiring the freeze-in condition to hold at all temperatures below reheating bounds the reheating temperature from above:
\begin{equation}
    T_{\rm reheat}\ll\dfrac{1}{(\alpha_{s}\Nc)^{3}}\dfrac{f_{a}^{2}}{M_{\rm pl}}~~.
    \label{eqn:reheat_bound_1}
\end{equation}
Below this temperature, we can ignore the $n_{\rm alp}/n_{\rm alp}^{\rm eq}$ ratio in equation \eqref{eqn:boltzmann_alp} and write it as
\begin{equation}
     -T\dfrac{dn_{\rm alp}}{dT}+3n_{\rm alp}=\dfrac{\Gamma_{gg\rightarrow ga}}{H}
\end{equation}
where $\dot{T}\approx -HT$. With the initial condition $n_{\rm alp}(T_{\rm reheat})=0$, the solution is
\begin{equation*}
    n_{\rm alp}(T)\sim \dfrac{M_{\rm pl}}{f_{a}^{2}} (T_{\rm reheat}-T)T^{3}~~.
    \label{eqn:freeze_in_density}
\end{equation*}
(Here we have dropped $\mathcal{O}(1)$ coefficients, and taken $\alpha_{s}\Nc\sim \mathcal{O}(1)$). In a Hubble patch, the number of particles thermally produced at $T\approx \Lambda$ is then 
\begin{equation}
    N_{\rm FI}\sim \dfrac{M_{\rm pl}}{f_{a}^{2}} (T_{\rm reheat}-\Lambda)\Lambda^{3}\dfrac{4\pi}{3 H_{\rm n}^{3}}~~.   
\end{equation}
We compare this to the number of particles produced by bubble expansion, $\langle N\rangle$, in the two scenarios discussed earlier. Requiring $N_{\rm FI}/\langle N\rangle < 0.1$, so that the freeze-in population is subdominant to that produced during expansion, the upper bound on the reheating temperature is the smaller of this bound and that of equation \eqref{eqn:reheat_bound_1}:
\begin{equation}
    T_{\rm reheat}^{\rm bound}\simeq\min\left\{\dfrac{f_{a}^{2}}{M_{\rm pl}}~,~\Lambda\left(1+0.1\left(\dfrac{\betagw}{H_{\rm n}}\right)^{3}\dfrac{\Lambda^{10}\gammaeff^2}{M_{\rm pl}^{4}m_{a}^{6}}\right)\right\}~,~~\gammaeff=\min\left\{\gammaterm,\dfrac{m_{a}}{\betagw}\right\}~~.
\end{equation}
It is straightforward to verify that for a large part of parameter space no fine-tuning of $T_{\rm reheat}$ is required. We now turn to the spectrum of the axions produced during expansion.

\subsection{Dark radiation and dark matter from the spectrum}
\label{subsec:spectrum_analysis}
\begin{figure}[t!]
    \centering
    \includegraphics[width=0.55\linewidth]{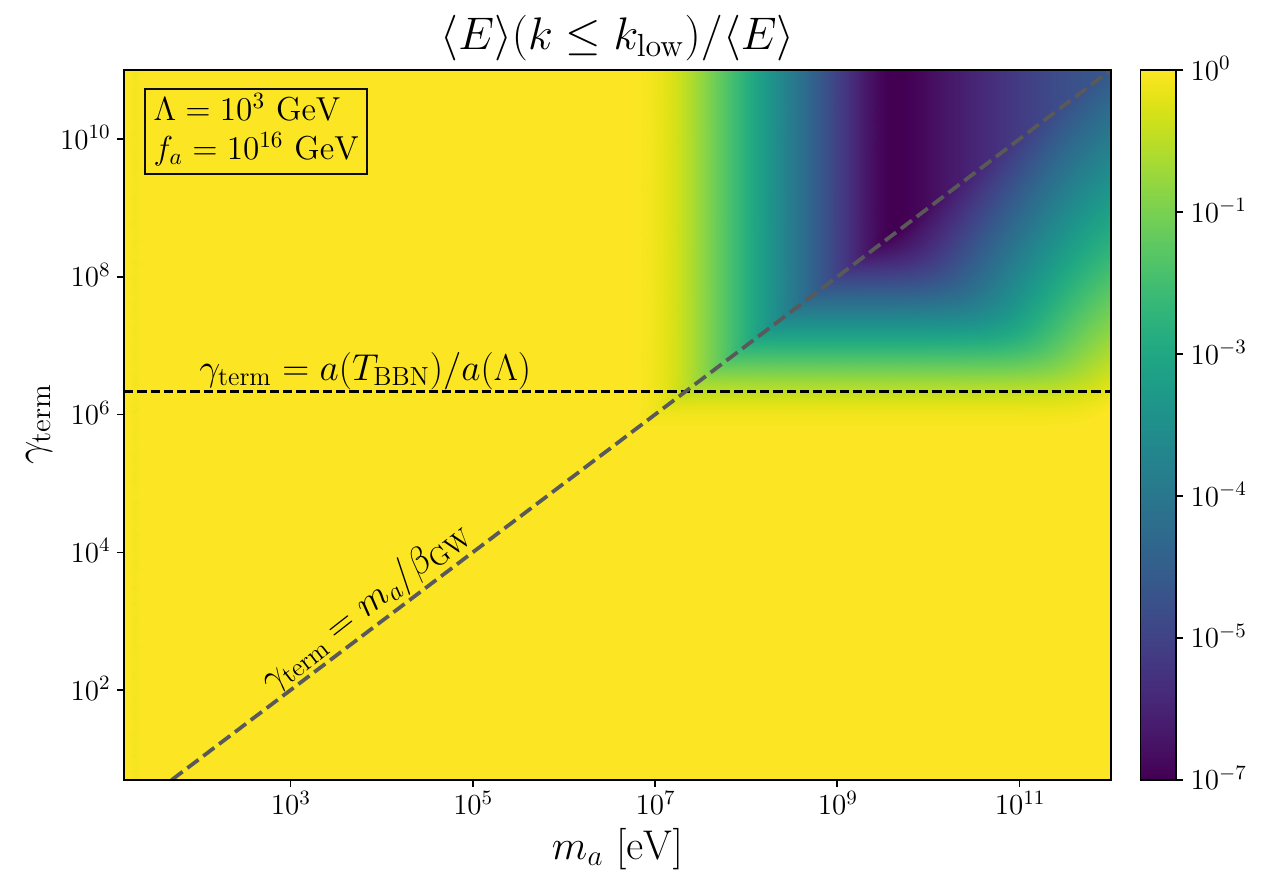}
    \caption{Fraction of energy in the low-energy modes $k\in[0,k_{\rm low}]$ in the terminal velocity case for $\{\Lambda=10^{3}~\text{GeV},f_{a}=10^{16}~\text{GeV},~g^{*}_{\rm s}(\Lambda)/g^{*}_{\rm s}(\tbbn)=10\}$. If $m_{a}$ is fixed $f_{a}$ only controls the coupling strength, and the fraction is independent of $f_{a}$. For $\gammaterm<a(\tbbn)/a(\Lambda)$, as the peak of the distribution $\sim \gammaterm m_{a}<k_{\rm low}$, most of the energy is in modes that are non-relativistic by BBN. For $\gammaterm>a(\tbbn)/a(\Lambda)$, as the peak is greater than $k_{\rm low}$, most of the energy is relativistic and hence the fraction is much smaller than one. For a fixed $\gammaterm$, larger $m_{a}$ leads to a larger relativistic fraction, but to a much smaller energy produced overall.}
    \label{fig:fraction_in_low_energy_modes}
\end{figure}

\begin{figure}[t!]
\begin{subfigure}[t]{0.49\linewidth}
 \centering  \includegraphics[width=\linewidth,height=5.5cm]{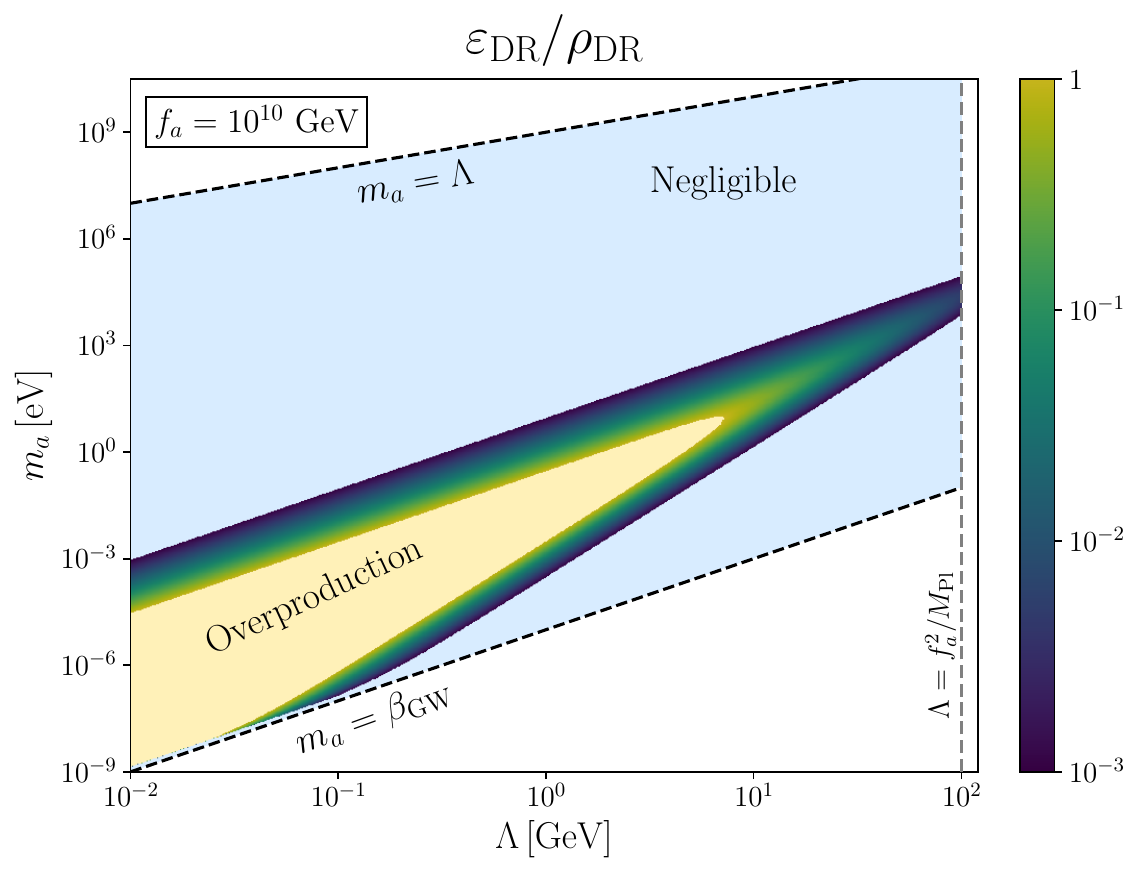}  
\caption{Ratio, in the runaway scenario, of the energy densities of the relativistic modes, $\epsdr$, and that of DR $\rhodr$ at temperature $\tbbn$ with $\Delta N_{\rm eff}=0.2$ at the current bound. We take $\betagw<m_{a}<\Lambda$. Light blue is the negligible region (ratio is $<10^{-3}$), while the light yellow region is excluded (ratio is $>1$). Parameters used: $\{f_{a}=10^{10}~\text{GeV},~\betagw/H_{\rm n}=10^{4},~\Delta N_{\rm eff}=0.2,~ g^{*}_{\rm s}(\Lambda)/g^{*}_{\rm s}(\tbbn)=10\}$.} 
      \label{fig:dark_radiation_runaway}
\end{subfigure}
\hspace*{0.2cm}
\begin{subfigure}[t]{0.49\linewidth}
   \centering
\includegraphics[width=\linewidth,height=5.5cm]{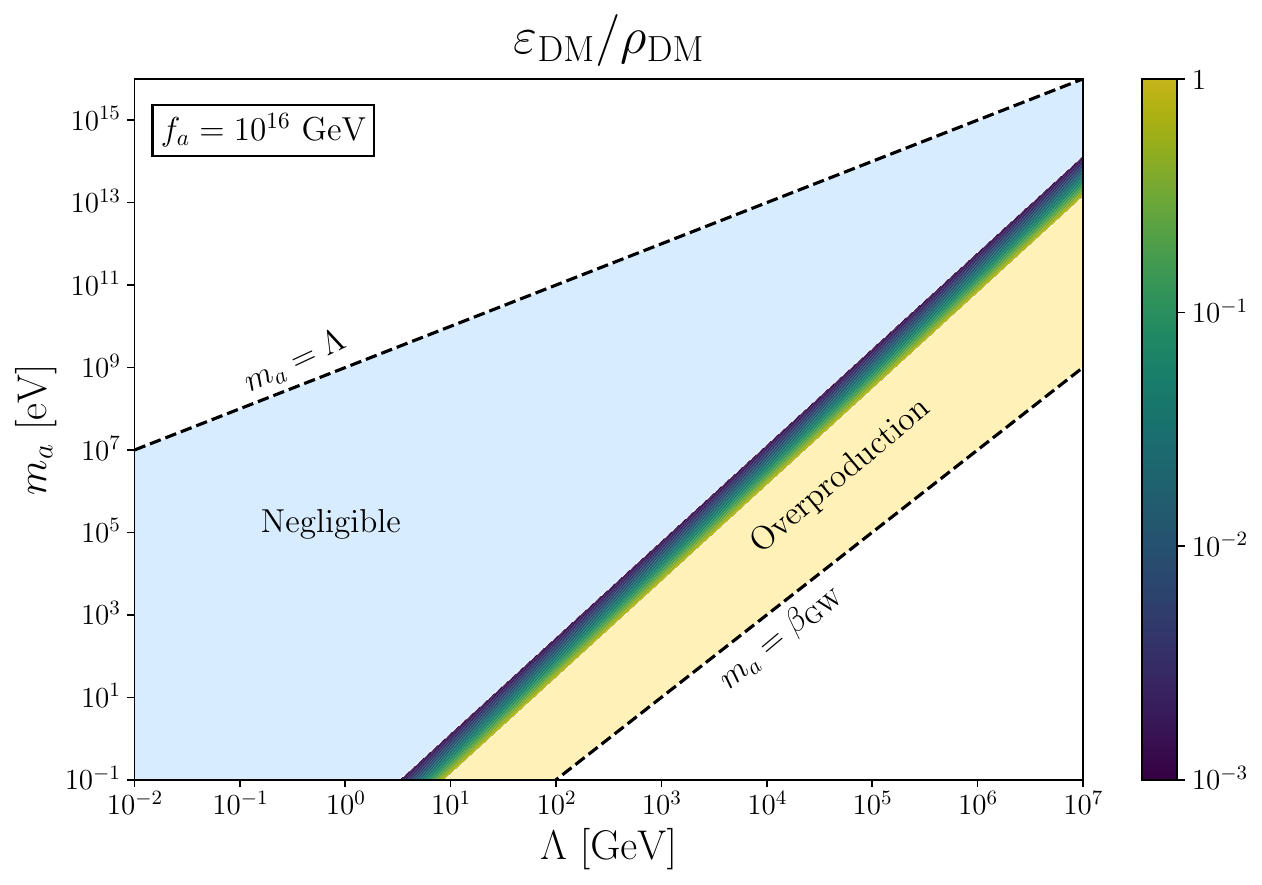}
\caption{Ratio of the energy density of the non-relativistic modes, $\epsdm$, redshifted at $\tbbn$ and the DM energy density, $\rhodm$, at BBN in the runaway scenario. We assume $\betagw<m_{a}<\Lambda$. Light blue is the negligible region (ratio is $<10^{-3}$), while the light yellow region is excluded (ratio is $>1$). Parameters used: $\{f_{a}=10^{16}~\text{GeV},~\betagw/H_{\rm n}=10^{4},~g^{*}_{\rm s}(\Lambda)/g^{*}_{\rm s}(\tbbn)=10\}$.} 
 \label{fig:dark_matter_runaway}
\end{subfigure}
\caption{Dark radiation and matter constraints in the runaway scenario.}
\end{figure}

\begin{figure}
   \begin{subfigure}[t]{0.49\linewidth}
          \centering
\includegraphics[width=\linewidth, height=5.5cm]{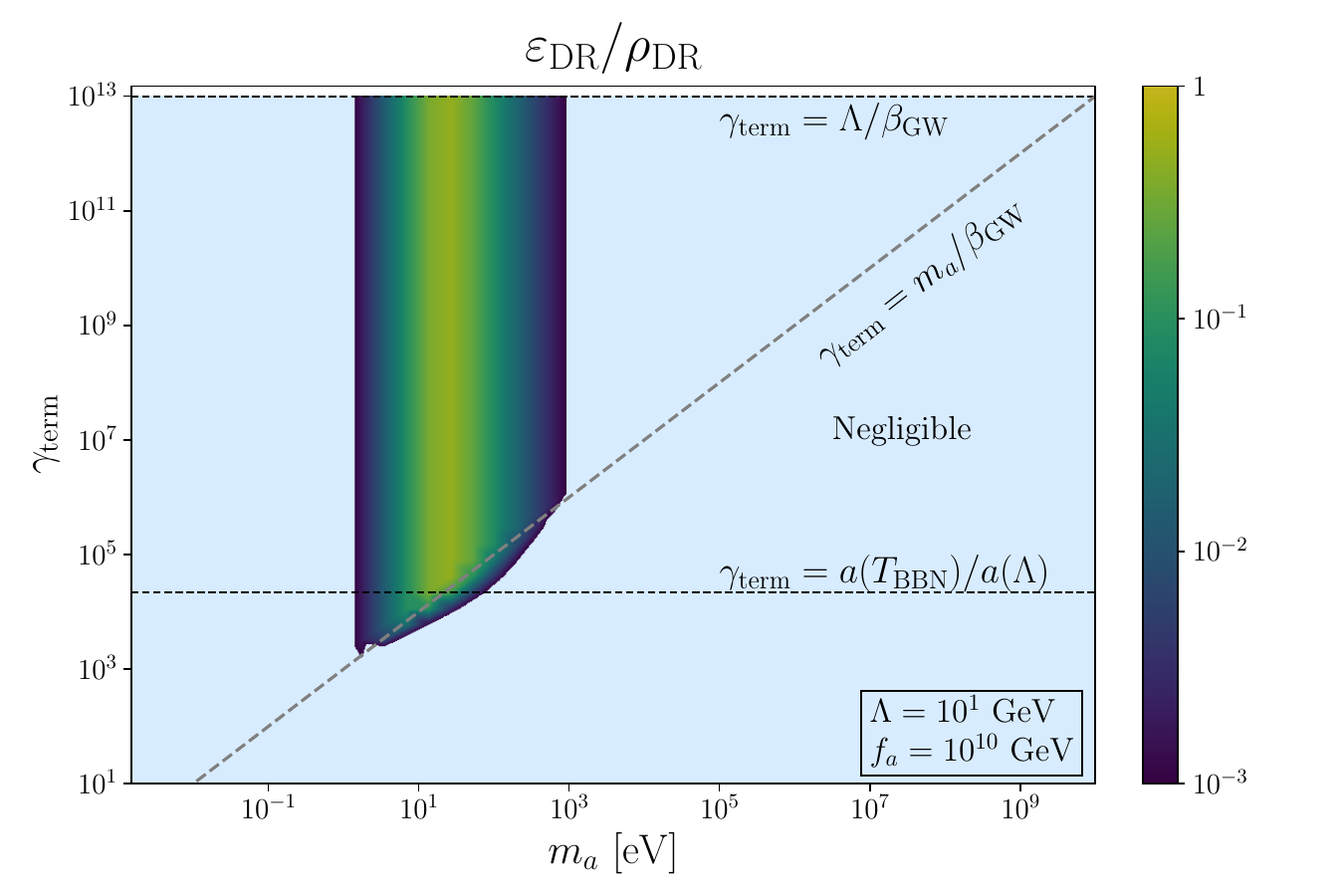}
 \caption{Ratio of the energy density of the relativistic modes $\epsdr$ and the energy density in the DR $\rhodr$ (with $\Delta N_{\rm eff}=0.2$) at temperature $\tbbn$ in the terminal velocity scenario. We assume $\betagw<m_{a}<\Lambda$. Light blue is the negligible region (ratio is $<10^{-3}$), while the light yellow region is excluded (ratio is $>1$). Parameters used: $\{f_{a}=10^{10}~\text{GeV},~\betagw/H_{\rm n}=10^{4},~\Delta N_{\rm eff}=0.2,~g^{*}_{\rm s}(\Lambda)/g^{*}_{\rm s}(\tbbn)=10\}$.} 
  \label{fig:dark_radiation_cst_vel}
   \end{subfigure}
   \hspace*{0.2cm}
    \begin{subfigure}[t]{0.49\linewidth}
          \centering
\includegraphics[width=\linewidth, height=5.5cm]{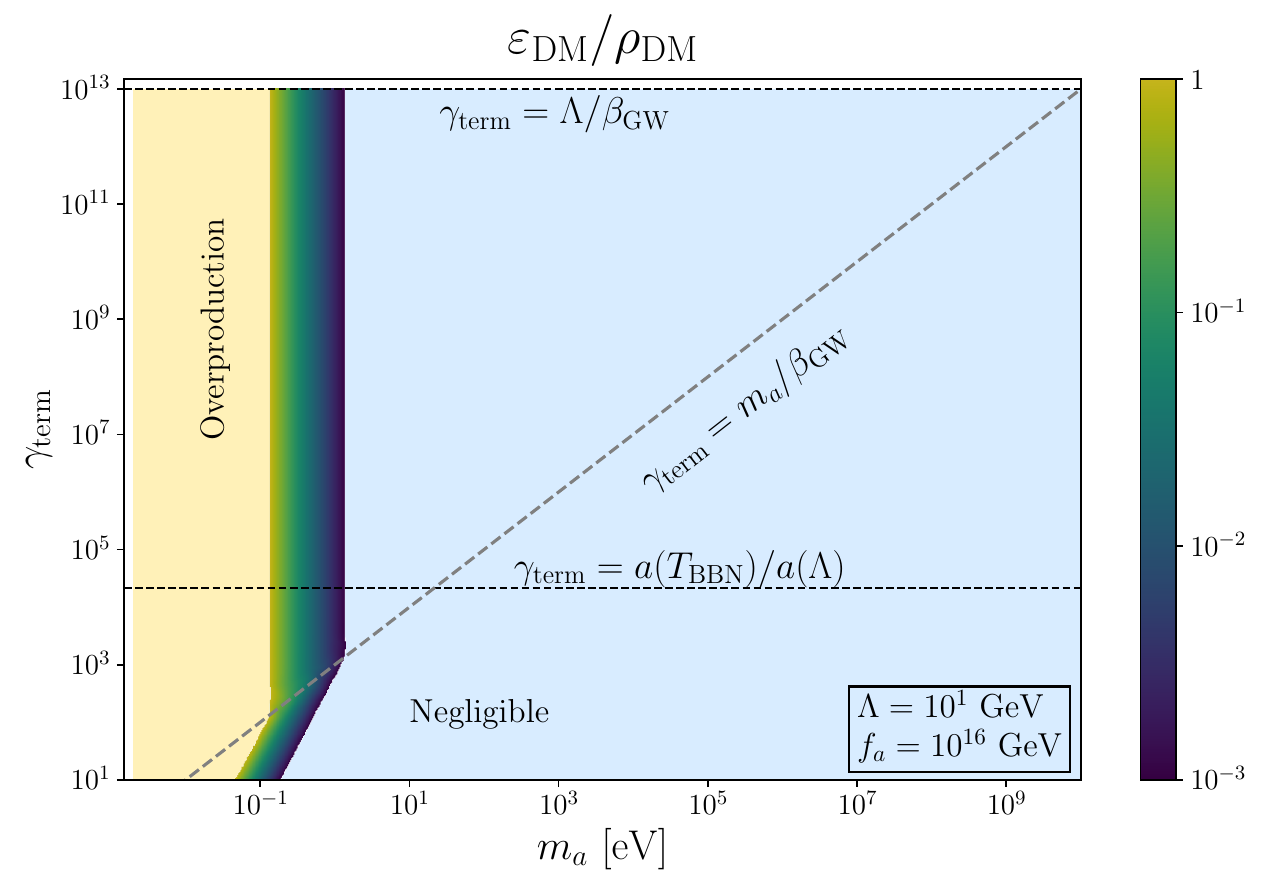} 
\caption{Ratio of the energy density of the non-relativistic modes $\epsdm$ redshifted to $\tbbn$ and the DM energy density $\rhodm$ at BBN in the terminal velocity scenario. We assume $\betagw<m_{a}<\Lambda$. Light blue is the negligible region (ratio is $<10^{-3}$), while the light yellow region is excluded (ratio is $>1$).  Parameters used: $\{f_{a}=10^{16}~\text{GeV},~\betagw/H_{\rm n}=10^{4},~g^{*}_{\rm s}(\Lambda)/g^{*}_{\rm s}(\tbbn)=10\}$.} 
 \label{fig:dark_matter_cst_vel}
\end{subfigure}
\caption{Dark radiation and matter constraints in $\{ m_a,\gammaterm\}$ space at $\Lambda=10^{1}~\text{GeV}$.}
\end{figure}

It is natural to consider the possibility that the particles produced by this mechanism can contribute to the dark matter (DM) or dark radiation (DR) densities.
To determine whether the produced energy density remains relativistic at a later cosmological epoch --- we focus on Big Bang Nucleosynthesis (BBN), with $\tbbn\approx 1~\text{MeV}$ --- we need the minimum momentum $k_{\rm low}$ at production (at $T\sim \Lambda$) such that a particle remains relativistic after redshifting to that epoch. This gives the condition
\begin{equation*}
    k>k_{\rm low}=\dfrac{a(\tbbn)}{a(\Lambda)}m_{a}~,~~\dfrac{a(\tbbn)}{a(\Lambda)}=\left(\dfrac{g_{\rm s}^{*}(\Lambda)}{g_{\rm s}^{*}(\tbbn)}\right)^{1/3}\dfrac{\Lambda}{\tbbn}~~.
    \label{eqn:k_min_runaway}
\end{equation*}
where $g^{*}_{\rm s}(T)$ is the effective number of degrees of freedom at temperature $T$.  As discussed in section \ref{sec:finite_wall_thickness}, there is also an upper bound on the produced momenta, set by the Lorentz-contracted thickness of the bubble wall:
\begin{equation}
    k< k_{\rm max}=\gammaterm \Lambda~~.
    \label{eqn:k_max_runaway}
\end{equation}
If $k_{\rm low}>k_{\rm max}$, essentially the entire spectrum (up to the exponentially suppressed tail) redshifts to non-relativistic momenta and contributes only to the DM abundance. By convention, we count the part of the spectrum below $k_{\rm low}$ as DM at $\tbbn$, and the interval $[k_{\rm low}, k_{\rm max}]$ as DR. If the bubble reaches a terminal velocity then, as discussed previously, most production occurs during the constant-velocity phase, since for $m_{a} R_0<1$ the radius at the end of the accelerated phase, $(R_0^2+\tauacc^2)^{1/2}$, is always smaller than $\gammaterm/m_{a}$. The peak of the spectrum scales as
\begin{equation}
    k_{\rm peak}\sim m_{a}\gammaeff~,~~~\gammaeff=\text{min}\left\{\gammaterm,\dfrac{m_{a}}{\betagw}\right\}~~.
\end{equation}
 Notice that since we have imposed that $m_a<\Lambda$ the peak will never cross into the region suppressed by the wall thickness. Thus, at the parametric level,
\begin{align*}
  k_{\rm peak} <k_{\rm low}\Rightarrow\gammaeff&<\dfrac{a(\tbbn)}{a(\Lambda)}\implies \text{dark matter}~~,\\
   k_{\rm peak}>k_{\rm low}\Rightarrow  \gammaeff&>\dfrac{a(\tbbn)}{a(\Lambda)}\implies \text{dark radiation}~~.
\end{align*}
This can be seen in figure \ref{fig:fraction_in_low_energy_modes}. 
At leading order, using equation \eqref{eqn:final_energy_scaling} and dividing by the bubble volume $\betagw^{-3}$ just before collision, the energy density is
\begin{equation}
    \varepsilon\sim  \frac{\betagw^3J_0^2\gammaeff^{3}}{m_a^5}~~.
\end{equation}
The energy density in the modes that are non-relativistic at $\tbbn$ is given by
\begin{equation}
    \epsdm=\left(\dfrac{a(\Lambda)}{a(\tbbn)}\right)^{3}\betagw^{3}\int_{0}^{k_{\rm low}}dk~\dfrac{d\langle E\rangle}{dk}\dfrac{\sqrt{m_{a}^{2}+(a(\Lambda)/a(\tbbn))^{2}k^{2}}}{\sqrt{m_{a}^{2}+k^{2}}}~~,
    \label{eqn:epsdm}
\end{equation}
where the factor inside the integral comes from the mode-by-mode redshift and the factor $(a(\Lambda)/a(\tbbn))^{3}$ outside is associated with the volume dilution due to expansion of the universe. We then compare this to the DM density at $\tbbn$, denoted by $\rhodm\approx 10^{-18}~\text{GeV}^{4}$ (see Figures \ref{fig:dark_matter_runaway} and \ref{fig:dark_matter_cst_vel}). For a given choice of $f_{a}$, this constrains the parameter space and identifies regions where the model contributes a significant fraction of the DM abundance.
           
Similarly, we compare the relativistic energy density $\epsdr$ with the allowed DR density at temperature $\tbbn$. Using a mode-by-mode redshift as in equation \eqref{eqn:epsdm}, we obtain
\begin{align}
    \epsdr&=\left(\dfrac{a(\Lambda)}{a(\tbbn)}\right)^{3}\betagw^{3}\int_{k_{\rm low}}^{k_{\rm max}}dk~\dfrac{d\langle E\rangle}{dk}\dfrac{\sqrt{m_{a}^{2}+(a(\Lambda)/a(\tbbn))^{2}k^{2}}}{\sqrt{m_{a}^{2}+k^{2}}}\nonumber\\
    &\approx  \left(\dfrac{a(\Lambda)}{a(\tbbn)}\right)^{4}\betagw^{3}\int_{k_{\rm low}}^{k_{\rm max}}dk~\dfrac{d\langle E\rangle}{dk}~~,
    \label{eqn:epsdr} 
\end{align}
where the approximation in the second line is valid when $k_{\rm peak}\gg k_{\rm low}$. This is the expected result: it is simply the dilution of energy density in the modes that remain relativistic throughout the cosmological evolution between $\Lambda$ and $\tbbn$. This is to be compared with the DR density bound at $\tbbn$, expressed in terms of one effective neutrino species:
\begin{equation}
    \rhodr=\Delta N_{\rm eff}\dfrac{\pi^{2}}{15}\dfrac{7}{8}\left(\dfrac{4}{11}\right)^{4/3}\tbbn^{4}~,~{\rm with}~\Delta N_{\rm eff}\approx 0.2~~.
    \label{eqn:rho_dr}
\end{equation}
Figures \ref{fig:dark_radiation_runaway} and \ref{fig:dark_radiation_cst_vel} show, for different scenarios as discussed below, the ratio $\epsdr/\rhodr$, which, if greater than one, corresponds to overproduction of DR compared to the current bounds, while $\epsdr/\rhodr < 1$ corresponds to a fraction of the current DR density bound at $\tbbn$. 

\subsubsection*{Benchmark plots}

Turning to the numerical evaluation of these constraints, we take two benchmark points with $\betagw/H_{\rm n}=10^{4}$ and $f_{a}\in\{10^{10},10^{16}\}~\text{GeV}$ and show that bubble expansion can yield a considerable DM or DR abundance across different parameter choices. For the constraints coming from DR we plot the ratio $\epsdr/\rhodr$ at $\tbbn$, where $\rhodr$ is defined in equation \eqref{eqn:rho_dr}, while for the DM we plot the ratio $\epsdm/\rhodm$ at $\tbbn$, where the redshifted $\epsdm$ is obtained using equation \eqref{eqn:epsdm}.

The free parameters are the axion mass $m_a$, the nucleation temperature $\Lambda$, and the bubble Lorentz factor $\gammaterm$ (equivalently, the acceleration time $\tauacc$). We consider two slices of this space: in the runaway case we vary $m_a$ and $\Lambda$, setting $\tauacc=\betagw^{-1}$ (itself a known function of $\Lambda$); in the terminal-velocity case we fix $\Lambda=10~\text{GeV}$ and vary $m_a$ and $\gammaterm$. In all plots, light blue indicates a negligible ratio, yellow denotes overproduction, and we assume $\betagw<m_{a}<\Lambda$, with $f_{a}=10^{16}~\text{GeV}$ for DM and $f_{a}=10^{10}~\text{GeV}$ for DR.
 \subsubsection*{Runaway scenario}
    \begin{enumerate}
        \item \textbf{Dark radiation} (figure \ref{fig:dark_radiation_runaway}): In the bottom-left corner DR is overproduced: the initial spectrum is highly relativistic and, with $\Lambda$ close to $\tbbn$, remains so ($k_{\rm low}$ is close to $m_{a}$). Away from $\tbbn$, for a fixed $\Lambda$ (which automatically also fixes $\betagw$) the peak of very light particles falls to the left of $k_{\rm low}$, and hence there is only a small fraction of relativistic modes. As the mass increases, the peak moves right towards and crosses $k_{\rm low}$ and so the fraction in relativistic modes increases. Yet further increasing the mass again leads to a negligible ratio because, although the relativistic fraction of total produced energy increases, the total energy decreases, scaling as $1/m_{a}^{2}$. The line $\Lambda=f_{a}^{2}/M_{\rm pl}$ marks $\Lambda=T_{\rm reheat}^{\rm bound}$; the transition is relevant only for scales below this value.

         \item \textbf{Dark matter} (figure \ref{fig:dark_matter_runaway}): In contrast to DR, production is negligible in the left corner of the plot, at light $m_{a}$ and $\Lambda$ close to $\tbbn$. This is expected as most of the energy is in relativistic modes. Away from $\tbbn$, lighter masses even lead to overproduction of DM. As $m_{a}$ increases, the peak moves to the other side of $k_{\rm low}$ while the energy produced decreases, so the ratio becomes negligible.
    \end{enumerate}
  \subsubsection*{Terminal Velocity Scenario}
    \begin{enumerate}
    \item \textbf{Dark Radiation} (figure \ref{fig:dark_radiation_cst_vel}): As discussed in section \ref{sec:hard_cutoff}, the spectrum saturates as $\gammaterm$ reaches $m_{a}/\betagw$ and so, for a fixed value of $m_{a}$, the region above the $\gammaterm=m_{a}/\betagw$ line has $\gammaterm$-independent behaviour. When $\gammaterm<a(\tbbn)/a(\Lambda)$, most of the energy is in non-relativistic modes. This means that in the parameter region $m_{a}/\betagw<a(\tbbn)/a(\Lambda)$, the majority of the energy is non-relativistic. Only for $m_{a}/\betagw\sim a(\tbbn)/a(\Lambda)$, when $k_{\rm peak}\sim k_{\rm low}$, is there a considerable fraction of energy going to relativistic modes. As $m_{a}$ increases, the story parallels the runaway case: although $k_{\rm peak}>k_{\rm low}$, so that the relativistic fraction approaches one, the total energy produced is suppressed by the growing mass. This is seen in the figure: as $m_a$ increases the ratio again becomes negligible. 

    \item \textbf{Dark Matter} (figure \ref{fig:dark_matter_cst_vel}): For $\gammaeff<a(\tbbn)/a(\Lambda)$, the spectrum redshifts into a predominantly DM contribution. As the energy in the axions is inversely proportional to $m_a$, smaller $m_a$ tends to overproduce DM. When $\gammaeff> a(\tbbn)/a(\Lambda)$, so that $k_{\rm peak}>k_{\rm low}$, the energy is mainly in relativistic modes and the DM contribution is negligible.
    \end{enumerate}

These ratios depend on $f_{a}$, so the regions of over- and underproduction shift accordingly. Nevertheless, these examples show that for a given value of $f_{a}$ one can obtain either DM, DR or both in some part of the $\{\Lambda, m_{a},\gammaterm\}$ parameter space. \footnote{For simplicity we considered the situation where the temperature of the dark sector was the same as that of the visible sector, but this is not necessary. This would quantitatively change the analysis but qualitatively the physics of particle production during the expansion phase would remain the same.}
\subsection{Backreaction from axion production}
\label{sec:back-reaction}
The explicit model also allows us to estimate whether, parametrically, back-reaction from particle production can, on its own, drive the bubble wall to a terminal velocity. We can estimate the effective back-reaction pressure on the wall using the basic expression
\begin{equation}
    p_{\rm rad}(t) = \frac{1}{4\pi R^2(t)} \frac{d\langle E\left(R(t)\right)\rangle}{dR}~.
\end{equation}
During the incoherent regime, $1/m \lesssim R \lesssim \gamma/m$, we have
\begin{equation}
    \langle E(R)\rangle \simeq v^6 \frac{J_0^2 R^3}{m^2}~\implies~ p_{\rm rad} \simeq v^6 \frac{J_0^2}{m^2}~.
\end{equation}
This $v$-dependent back-reaction pressure can lead to a terminal-velocity epoch if
\begin{equation}
    p_{\rm rad} \gtrsim \Delta P~,
\label{eqn:back-reaction-condition}
\end{equation}
where $\Delta P$ is the pressure difference across the walls in the \emph{absence} of axion production.  

For the dynamics of bubble walls in a thermal PT in the quasi-equilibrium regime (so wall motion is not so fast that shocks and other far-from-equilibrium phenomena occur), $\Delta P = \Delta F$, the free-energy difference between the false and true vacua. Importantly, $\Delta F$ must be evaluated \emph{at the bubble nucleation temperature} $\tnuc < \tcr$ (see appendix \ref{appendix:transition_dynamics}) and possibly somewhat below, and not at $T=0$ where $\Delta F$ would be the full, large latent heat of the transition.  For the $SU(\Nc)$ YM-axion model we expect, parametrically, that 
\begin{equation}
    \Delta F \sim \epsilon(\tnuc/\tcr) \Nc^2 \Lambda^4~.
\end{equation}
Here, the possibly very small numerical factor $\epsilon(\tnuc/\tcr)$ accounts for the fact that strongly coupled FOPTs are often barely supercooled, $(1- \tnuc/\tcr)\ll 1$, so that the free-energy difference across a wall in a realistic $SU(\Nc)$ transition can be parametrically much smaller than the latent heat of the transition. \footnote{This large suppression in the relevant pressure difference across the deconfinement-confinement wall occurs for pure $SU(\Nc)$ theories at $\theta=0$ with potentially very significant consequences for PT dynamics (and gravitational wave signatures).  For a discussion of this under-appreciated fact see, e.g., \cite{Lucini:2005vg,GarciaGarcia:2015fol,Gouttenoire:2021kjv,Gouttenoire:2023roe,Agrawal:2025wvf,Agrawal:2025xul}. The $\theta_{\rm eff}$ dependence of $\epsilon(\tnuc/\tcr)$ is not known; we here assume it is an additional $\mathcal{O}(1)$ factor which we suppress. We thus take $\epsilon(\tnuc/\tcr)$ to be parametrically smaller than unity.} In our axion theory, assuming that $\muqp^2 > \Lambda^2$ and that $\delta \sim \mathcal{O}(1)$ (which is natural in this regime), we have $J_0 \sim \Lambda^4/ f_a$, $m_a \sim \muqp^2 /f_a$ and thus 
\begin{equation}
    p_{\rm rad} \sim v^6 \frac{\Lambda^8}{\muqp^4}~.
\end{equation}
The condition in equation \eqref{eqn:back-reaction-condition} for significant back-reaction is therefore
\begin{equation}
    v^6 \left(\frac{\Lambda}{\muqp}\right)^4 \gtrsim \epsilon(\tnuc/\tcr) \Nc^2 ~.
 \label{eqn:axion-back-reaction}
\end{equation}
Many factors in this expression are uncertain, but if the PQ-violation scale $\muqp\not\gg \Lambda$, back-reaction can be important. The most interesting possibility is that axion production alone drives a terminal-velocity epoch at quasi-relativistic speeds, $v\sim 0.1-0.3$; this requires both $\muqp\sim \Lambda$ and $\epsilon(\tnuc/\tcr)\ll 1$. Note that in the case $\muqp^2 \ll \Lambda^2$ the additional PQ-violating potential is a small perturbation to the $SU(\Nc)$-generated potential, and the axion relaxes the effective $\theta$-term to small values, $\theta_{\rm eff}\ll 1$. In this case $p_{\rm rad}$ is suppressed by a factor of $\sin^2(\theta_{\rm eff})$, further reducing the back-reaction pressure from axion production.

\section{Conclusions}
\label{sec:discussion}

The central result of this work is that the expansion of true-vacuum bubbles during a first-order phase transition can, by itself, be a copious source of light spin-0 particles. For a field linearly coupled to the bubble profile, we showed that production persists even for walls that expand at constant velocity --- contrary to the common expectation that a steadily moving wall cannot radiate. Production becomes inefficient only once the local-rest-frame radius of curvature of the wall exceeds the particle Compton wavelength, corresponding to a FRW-frame bubble radius $R\sim\gamma/m$. For light particles this epoch can be long-lasting, and the resulting population can parametrically dominate other production mechanisms of feebly coupled particles, including freeze-in. Along the way, we computed the momentum spectrum for constant-velocity, runaway, and patched trajectories; identified the Compton wavelength as the scale separating coherent production, scaling as the square of the wall area, from patch-coherent production, scaling as the area itself; quantified the weak-source condition under which our perturbative treatment is valid; and estimated the effects of finite wall thickness.

We applied these results to a hidden-sector $SU(\Nc)$ Yang-Mills theory, with $\Nc\geq 3$, undergoing a thermal deconfinement-confinement transition in the presence of an axion-like particle with an imperfectly protected shift symmetry. Over large regions of parameter space, the ALPs produced during bubble expansion dominate the freeze-in population and can account for a significant fraction of dark radiation and/or dark matter, with warm dark matter in a transition region. Since we have not included production during bubble collisions (a computation requiring dedicated numerical simulation of the multi-bubble field dynamics), our estimates constitute a \emph{lower bound} on the total yield.

Several directions merit further investigation. In the confining case, the post-transition phenomenology is particularly rich: the presence of the axion, together with the violation of P and CP, opens new decay channels for glueball states that would otherwise be stabilized by their $J^{PC}$ quantum numbers \cite{Morningstar:1999rf}, potentially yielding a combination of heavy and light dark matter states. This particle production mechanism should also leave an imprint on the gravitational wave signal of the transition, both because particle production drains kinetic energy from the walls and because the produced particles can themselves source gravitational waves \cite{Inomata:2024rkt,Ghoshal:2026hev},\footnote{We thank Anish Ghoshal for discussions on this interesting direction of investigation.} and quantifying these effects requires a dedicated analysis. Finally, if the electroweak phase transition is rendered first-order by new physics, this mechanism applies to ultralight scalars coupled to the Higgs sector \cite{Piazza:2010ye}, providing new bounds on their masses and couplings --- a question we plan to return to in future work.

\section*{Acknowledgements}
The research of IGG and AP has been supported by the U.S.~Department of Energy grant No.~DE-SC0011637. GRK expresses gratitude to Somerville College, Oxford for support via the Oxford Ryniker-Lloyd Graduate Scholarship jointly with a Clarendon Fund Scholarship. JMR thanks the University of Washington theory group for hospitality during a portion of this work. The authors acknowledge the use of OpenAI Codex for assistance with code development and polishing prose during the preparation of this work. All results, analysis, and conclusions were verified by the authors. For the purpose of Open Access, the authors have applied a CC BY public copyright licence to any Author Accepted Manuscript version arising from this submission.

\appendix

\section{Condition for weak coupling during coherent production}
\label{appendix:weak_coupling}
We now discuss the condition ensuring that the $|{\tilde J}|^2$ terms dominate the production rate.  This is an issue even if the source is very weakly coupled as, in the presence of $\varphi$ self-interactions and during epochs of \emph{coherently enhanced} production, the source can effectively behave as if it were strong. As we saw in section \ref{subsec:constant_velocity_and_the_regulator}, particle production is coherent until the bubble radius $R$ reaches $m^{-1}$, so we need to find a ``weak source'' condition in this case.\footnote{Our discussion is motivated by that of reference \cite{Gelis:2006yv} on particle production due to strong sources.} 

The connected generating functional $W[J]$ for a source $J$ linearly coupled to $\varphi$ is
\begin{equation}
   e^{iW[J]}=\int [D\varphi]~~e^{i(S[\varphi]+\int d^{4}x~J(x)\varphi(x))}~~.
\end{equation}
Equivalently, this is the transition amplitude $\langle 0_{\rm out}\vert 0_{\rm in}\rangle$ where $\ket{0_{\rm in}}$ and $\ket{0_{\rm out}}$ are the asymptotic past and future vacuum states. Hence the probability of particle production is
\begin{equation}
    P_{\rm prod}=1-\vert\langle 0_{\rm out}\vert 0_{\rm in}\rangle\vert^{2}=1-e^{-2\text{Im}(W[J])}~~.
\end{equation}
$W[J]$ can be expanded as (the source is $\otimes$)
\begin{equation}
    W[J]=~\feynmandiagram[inline=(b.base), horizontal=a to b]{a[crossed dot]--b[crossed dot],};~+\sum_{n\geq 3}~ \triplediagram[0.25]{n}+\cdots
     \label{eqn:expansion_of_generating_functional}
\end{equation}
where the second diagram is the tree-level $n$-point interaction term and $\cdots$ contains all other terms including loops.  Using the cutting rule to calculate
${\rm Im}(W[J])$, the first diagram gives the phase space integral of $\vert \widetilde{J}\vert^2 $ ($\widetilde{J}$ being the FT). The second diagram in equation \eqref{eqn:expansion_of_generating_functional} is
\begin{equation}
  \triplediagram[0.25]{n}=
    \dfrac{\lambda_{n}}{(2\pi)^{4n}} \left(\int \prod_{i=1}^{n} d^{4} p_{i}\right) \left[\prod_{j=1}^{n}\widetilde{J}(p_{j})\widetilde{D}(p_{j}) \delta^{(4)}\left(\sum_{i=1}^{n}p_{i}\right)\right]~~,
    \label{eqn:quartic_diagram}
\end{equation}
where $\lambda_{n}$ is the $n-$point vertex coupling. Consider a bubble of initial size $R_{0}\rightarrow 0$ (a valid limit in the case of most interest, $mR_{0}<1$) expanding with constant relativistic speed $v$, so
\begin{equation*}
    \widetilde{J}(\omega,\vec{k})=\dfrac{8\pi J_{0}v^{3}}{(\omega^{2}-\vert\vec{k}\vert^{2}v^{2})^{2}} ~~~.
\end{equation*}
A single cut on the external leg can be associated with a one-particle state \cite{Gelis:2006cr}:
\begin{align}
    \cuttriplediagram[0.25]{n}
    =\dfrac{\lambda_{n}(8\pi J_{0}v)^{n}}{(2\pi)^{4n-5}m^{2n+4}}\gamma^{2}\int \dfrac{d^{3}x}{\sqrt{x^{2}+\gamma^{-2}}}\dfrac{\mathcal{A}_{n}(\gamma x)}{(1+x^{2})^{2}}~~.
\end{align}
To extract its $\gamma$ scaling, consider $\mathcal{A}_{n}(Q)$, a smooth function that decays as $Q\rightarrow\infty$, 
\begin{equation}
   \vert \mathcal{A}_{n}(Q)\vert\leq \dfrac{C}{(1+\vert Q\vert)^{r}}~,~~r> 0~~,~~C\sim \mathcal{O}(1)~~.
\end{equation}
This implies that the integral is bounded as
\begin{equation*}
    \int \dfrac{d^{3}x}{\sqrt{x^{2}+\gamma^{-2}}}\dfrac{\mathcal{A}_{n}(\gamma x)}{(1+x^{2})^{2}}\lesssim \int \dfrac{d^{3}x}{\sqrt{x^{2}+\gamma^{-2}}}\dfrac{1}{(1+\vert x\vert\gamma)^{r}(1+x^{2})^{2}}\leq \int \dfrac{d^{3}x}{\vert x\vert (1+x^{2})^{2}}=2\pi~~.
\end{equation*}
One can verify that, as $\vert Q\vert \rightarrow \infty$, $\mathcal{A}_{n}\sim \vert Q\vert^{-2n}$ if all momenta $p_{i}$ on the uncut legs scale as $\vert Q\vert$, whereas $\mathcal{A}_{n}\sim \vert Q\vert^{-4}$ if they remain $\mathcal{O}(1)$. Hence, at most, 
\begin{equation}
\cuttriplediagram[0.25]{n}\lesssim \dfrac{2\lambda_{n}(J_{0}v)^{n}}{\pi^{2n-1} m^{2n+4}}\gamma^{2}    
\end{equation}
For this to be subdominant to the first term in equation \eqref{eqn:expansion_of_generating_functional}, we parametrically require
\begin{equation}
    \dfrac{\lambda_{n}J_{0}^{n-2}}{m^{2n-2}}\ll 1~~.
    \label{eq:weak_source_appendix}
\end{equation}
For the axion model of section \ref{sec:alp_pheno_case}, this is easily satisfied. 
\section{Details of \texorpdfstring{$SU(\Nc)$}{SU(Nc)}-axion model transition dynamics}
\label{appendix:transition_dynamics}

FOPT dynamics in strongly coupled theories is an active topic of study, in both its theoretical and observational aspects --- in particular gravitational wave generation (see \cite{Agrawal:2025wvf, Fujikura:2025iam} and references therein). Here we briefly recap some aspects of the $SU(\Nc)$ Yang-Mills theory case. At zero $\theta$-parameter and chemical potential, the thermal deconfinement-confinement transition is expected to have small supercooling \cite{Lucini:2005vg,GarciaGarcia:2015fol,Salami:2025iqq,Agrawal:2025wvf,Agrawal:2025xul}. The quantity that dictates the transition rate is the thermal Euclidean bounce action $S_{\rm b}$. In the thin-wall limit
\begin{equation}
    S_{\rm b}(T)=\frac{16\pi}{3}\dfrac{\sigma_{\rm BW}^{3}}{Q_{\rm h}^{2}\tcr}\left(1-\dfrac{T}{\tcr}\right)^{-2}~~,
\end{equation}
where $\tcr$ is the critical temperature, $\sigma_{\rm BW}$ is the bubble wall tension and $Q_{\rm h}$ is the PT latent heat. Using $SU(\Nc)$ lattice data in the large-$\Nc$ limit \cite{Lucini:2005vg,Salami:2025iqq}, at $T\sim \tcr$ the action is estimated as
\begin{equation}
    S_{\rm b}\approx 8.86\times 10^{-4} N^{2}_{\rm c}\left(1-\dfrac{10.05}{\Nc^{2}}\right)^{3}\left(1-\dfrac{T}{\tcr}\right)^{-2}~~.
\end{equation}
For the nucleation to become efficient, we need at least one bubble per Hubble time per Hubble patch. This gives a relation between the nucleation temperature and the bounce action\footnote{In strongly coupled theories, the critical temperature and the nucleation temperature, $\tnuc$,  are very close to each other and are of the order of the dynamical confinement scale $\Lambda$.} \cite{GarciaGarcia:2015fol,Asadi:2021pwo,Agrawal:2025xul}. For simplicity we consider the nucleation temperature to be $\Lambda$ as this will not affect any of our main results. $\tnuc\simeq\Lambda$ and $S_{\rm b}$
\begin{equation}
    H_{\rm n}^{4}\equiv H^{4}(\tnuc)\approx \Lambda^{4}e^{-S_{\rm b}}~~\Rightarrow~~ S_{\rm b}\approx 4\ln\left(\dfrac{M_{\rm pl}}{\Lambda}\right)~~.
\end{equation}
For large $\Nc$, we can write the inverse of the duration of the FOPT, $\betagw$, as
\begin{equation}
   \dfrac{\betagw}{H_{\rm n}}=T\dfrac{dS_{\rm b}}{dT}\Bigg\vert_{T=\Lambda}\approx\dfrac{0.67\times 10^{2}}{\Nc} \left(4\ln\left(\dfrac{M_{\rm pl}}{\Lambda}\right)\right)^{3/2}~~.
   \label{eqn:beta_gw_def}
\end{equation}
The above definition of $\betagw$ is valid in the thin-wall limit.  When the transition is strongly coupled and/or away from the thin-wall limit, which is usually the case for supercooled transitions, one needs to know the full form of the action in order to calculate $\betagw$ using the definition of the percolation fraction \cite{Hindmarsh:2019phv}. In our calculations, we take $\betagw/H_{\rm n}=10^{4}$.
\bibliography{references}
\bibliographystyle{JHEP}
\end{document}